\documentclass[%
 reprint,
 superscriptaddress, % => affiliation's address tries to be 1 line
 amsmath,amssymb,
 aps,
 prb, % for Phys. Rev. B
]{revtex4-2}

\usepackage{graphicx}% Include figure files
\usepackage{dcolumn}% Align table columns on decimal point
\usepackage{bm}% bold math
\usepackage{hyperref}
\usepackage{braket}

\usepackage{textcomp}%https://munibus.hatenablog.com/entry/2018/02/07/065938

\usepackage{mathcomp} %->error \dagger is already defined
\usepackage{amsmath,amsthm,amssymb,cases}

\usepackage{here}
\usepackage{url} % for http://~ hyperlink
\usepackage{physics}

\begin{document}

%\preprint{APS/123-QED}

\title{Self-consistent study of topological superconductivity in two-dimensional quasicrystals}% Force line breaks with \\

\author{Masahiro Hori}
\affiliation{%
 Department of Physics and Engineering Physics, and Centre for Quantum Topology and Its Applications (quanTA), University of Saskatchewan, 116 Science Place, Saskatoon, Saskatchewan, Canada S7N 5E2
}%
\affiliation{%
 Department of Applied Physics, Tokyo University of Science, Tokyo 125-8585, Japan
}%

\author{Takanori Sugimoto}
%\affiliation{%
% Department of Applied Physics, Tokyo University of Science, Tokyo 125-%8585, Japan
%}%
\affiliation{%
Center for Quantum Information and Quantum Biology, Osaka University, Osaka 560-0043, Japan
}%

\author{Takami Tohyama}
\affiliation{%
 Department of Applied Physics, Tokyo University of Science, Tokyo 125-8585, Japan
}%

\author{K. Tanaka}
\email{Contact author: kat221@usask.ca}
\affiliation{%
 Department of Physics and Engineering Physics, and Centre for Quantum Topology and Its Applications (quanTA), University of Saskatchewan, 116 Science Place, Saskatoon, Saskatchewan, Canada S7N 5E2
}%

\date{\today}% It is always \today, today,

\begin{abstract}
We study two-dimensional $s$-wave topological superconductivity with Rashba spin-orbit coupling and Zeeman field in Penrose and Ammann-Beenker quasicrystals. By solving the Bogoliubov-de Gennes equations self-consistently for not only the superconducting order parameter, but also the spin-dependent Hartree potential, we show the stable occurrence of  topological superconductivity with broken time-reversal symmetry in both Penrose and Ammann-Beenker quasicrystals.
The topological nature of the quasicrystalline system is signified by the Bott index $B$. Topological phase transitions are found to occur, where $B$ changes between 0 and $\pm 1$, as the chemical potential or Zeeman field is varied. In terms of self-consistent solutions, we demonstrate the existence of a Majorana zero mode per edge or vortex when $B=\pm 1$, consistently with the bulk-edge/defect correspondence for periodic systems.

%\begin{description}
%\item[DOI]
%%CHANGE HERE 
%00.0000/PhysRevB.00.000000
%\end{description}
\end{abstract}

\maketitle

%\tableofcontents

\section{\label{sec:introduction}INTRODUCTION}
There are growing appeals for quasicrystals (QCs), unconventional materials with quasiperiodic order and rotational symmetry that is forbidden by crystallography \cite{Lalena_2006}, from the fundamental as well as application points of view. Mathematically, a QC can be obtained 
as projection of a periodic lattice in higher-dimensional space (`hyperspace' or superspace) called a hypercubic lattice onto lower dimensions. 
It is this relation of quasiperiodicity to higher dimensions that
makes QCs all the more interesting, with additional degrees of freedom such as phasons and capability of hosting topological phenomena 
\cite{Kraus_2016,Zilberberg_2021,Fan_2022,Koshino_2022}.
While research on QCs has a long history since its first discovery in 1984 \cite{Shechtman_1984,Levine_1984,Steurer_2018}, QCs have recently been studied intensively \cite{Nagao_2015,Dotera_2017,Yoshida_2019,Chang_2021,Tamura_2021}. 
QCs have been realised not only in solid-state materials, but also in 
optical systems \cite{Guidoni_1997,Verbin_2013,Bandres_2016} and quasicrystalline monolayers assembled on substrates \cite{Ledieu_2004,Sharma_2005,Ledieu_2008,Smerdon_2008,Forster_2013,Collins_2017}. 
Superconductivity has been discovered in an icosahedral Al-Zn-Mg quasicrystal \cite{Kamiya_2018} and 
a van der Waals-layered Ta-Te quasicrystal \cite{Tokumoto_23}. Theoretically,
unconventional features of superconductivity in QCs have been discussed in terms of the attractive Hubbard model \cite{Sakai_2017,Araujo_2019,Takemori_2020} and its extension \cite{Sakai_2019}.
Moreover, recent theoretical studies have focused on QCs as topological materials.
For example, the integer quantum Hall effect \cite{Kraus_2013,Tran_2015,Fuchs_2018}, the quantum spin Hall effect \cite{Huang_2018,Huang_2018_2,Huang_2019}, and higher-order topological phases \cite{Varjas_2019,Chen_2020} have been explored. It has also been shown that QCs can host weak topological superconductivity phases, which can be characterised by a real-space weak topological invariant \cite{Fulga_2016}.

Two-dimensional (2D) topological superconductors are promising candidate materials for creating and manipulating Majorana zero modes for topological quantum computation \cite{Nayak_2008,Stern_2010,Alicea_2012,Ando_2015,Sato_2017,Sun_2017}.
In accordance with the bulk-edge \cite{Hatsugai_1993PRL,Hatsugai_1993PRB,Fukui_2012} or bulk-defect \cite{Teo_2010} correspondence, a Majorana zero mode can appear on a surface boundary or in a vortex of a topological superconductor. 
In particular, chiral Majorana edge modes have been detected along the surface of a 2D topological superconductor, a cobalt island covered by monolayer lead \cite{Menard_2017}. This is an experimental realisation of 2D topological superconductivity (TSC) with broken time-reversal symmetry, first proposed by Sato, Takahashi, and Fujimoto \cite{Sato_2009_3,Sato_2010} in terms of a tight-binding model that combines conventional $s$-wave superconductivity, Rashba spin-orbit (SO) coupling, and Zeeman field. 
This system belongs to class D \cite{Schnyder_2008}, with the topological invariant $\nu$ being the first Chern number or the Thouless-Kohmoto-Nightingale-Nijs (TKNN) number \cite{Thouless_1982}. As a solution to the Bogoliubov-de Gennes (BdG) equations \cite{deGennes_1999,Zhu_2016}, the existence of a Majorana zero mode in a vortex in non-Abelian phase has been shown analytically \cite{Sato_2009_3,Sato_2010,Sau_2010,Tewari_2010} as well as 
numerically in terms of self-consistent solutions for a single vortex in an open-boundary system \cite{Bjoernson_2013} and for the vortex lattice \cite{Smith_2016}.

Ghadimi \textit{et al}. \cite{Ghadimi_2021} have generalised the tight-binding model of Ref.~\cite{Sato_2009_3,Sato_2010} for quasicrystalline systems and applied it to two prototypical 2D quasicrystals (QCs), Penrose \cite{Penrose_1974} and Ammann-Beenker (AB) QCs \cite{Grunbaum_1986,Beenker_1982}. In the absence of periodicity that enables calculation of the first Chern number, 
one can identify topological states by means of calculating the Bott index $B$ in real space \cite{Loring_2010,Loring_2019}, which is equivalent to the first Chern number in the thermodynamic limit \cite{Ge_2017,Yang_2019}. 
In Ref.~\cite{Ghadimi_2021}, by assuming a real constant for the superconducting order parameter at all sites, it has been found in both QCs that phase transitions between topologically trivial ($B=0$) and nontrivial ($B=1$) phases can occur and a Majorana zero mode appears as a chiral edge state or vortex bound state when $B=1$.
Due to the underlying inhomogeneity of a quasicrystalline lattice, however, the assumption of a uniform (real) superconducting order parameter may be too crude an approximation. In order to examine the occurrence of TSC in QCs, it is crucial to solve for the mean fields self-consistently, to take the inhomogeneity of QCs fully into account.

In the present work, we examine the occurrence of TSC in Penrose and AB QCs by solving the BdG equations on the tight-binding model \cite{Sato_2009_3,Sato_2010,Ghadimi_2021} self-consistently for the superconducting order parameter as well as the spin-dependent Hartree potential. We calculate the Bott index in terms of self-consistent solutions to signify the topological nature of the system. As periodic boundary conditions are necessary for calculating the Bott index, we use approximants of Penrose and AB QCs (and henceforth call them QCs). We find that the converged mean fields are both spatially inhomogeneous, reflecting the underlying structure of a QC. 
Furthermore, we show in terms of self-consistent solutions the existence of a zero-energy Majorana bound state per edge or vortex when the Bott index $B=\pm 1$.

The paper is organised as follows. In Sec.~\ref{sec:model} Penrose and AB QCs and their representation in the perpendicular space -- the complementary space to 2D real space -- are introduced, and the BdG equations on the tight-binding model \cite{Sato_2009_3,Sato_2010,Ghadimi_2021} and calculation of the Bott index are described briefly.
Results of our self-consistent calculations are presented and discussed for Penrose and AB QCs in Sec.~\ref{sec:PenQCs} and \ref{sec:ABQCs}, respectively. In Sec.~\ref{sec:impurities} the effects of randomly distributed nonmagnetic impurities on TSC in QCs are studied.
The work is summarised and discussed further in Sec.~\ref{sec:conclusion}. In Appendix, topological phase transitions in a square lattice are illustrated in terms of self-consistent solutions for comparison with the analogous results for QCs.

\begin{figure}[tbh]
\includegraphics[width=8cm]{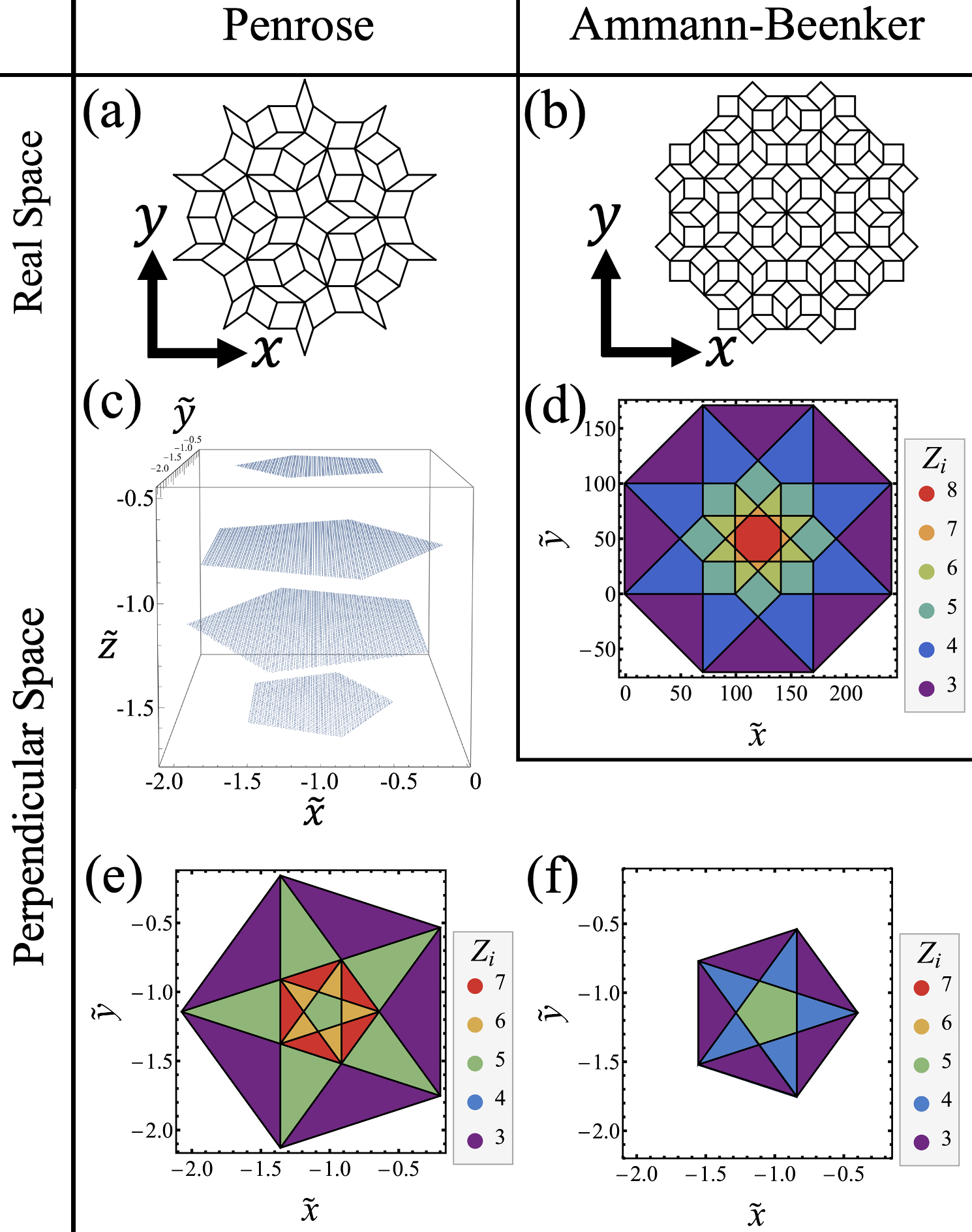}
\caption{(a) Penrose and (b) AB QCs in real space. (c) The three-dimensional perpendicular space of Penrose QCs, with four pentagonal planes at the $\tilde{z}$ coordinate of $-n/\sqrt{5}$, $n=1,2,3,4$.
(d) The two-dimensional perpendicular space of AB QCs, and the perpendicular space of Penrose QCs at (e) $\tilde{z}=-3/\sqrt{5}$ and (f) $\tilde{z}=-4/\sqrt{5}$, to which the pentagonal planes at $\tilde{z}=-2/\sqrt{5}$ and $\tilde{z}=-1/\sqrt{5}$ are identical, respectively, with rotation by $\pi/5$ and $-\pi/5$. In (d-f) lattice sites in real space with different values of the coordination number $Z_i$ are represented in different colours.
\label{fig:figPenAB}}
\end{figure}

\section{\label{sec:model}MODEL}

We solve the BdG equations on the tight-binding model for TSC \cite{Sato_2009_3,Sato_2010} generalised for Penrose and AB QCs \cite{Ghadimi_2021} self-consistently, for not only the superconducting order parameter, but also the spin-dependent Hartree potential. We use the vertex model, 
where the vertices of the quasiperiodic tiling are the lattice sites of a QC and the links connecting vertices are of the same length in both Penrose and AB QCs. 
Figure~\ref{fig:figPenAB} illustrates (a) Penrose and (b) AB QCs in real space ($xy$ plane), which are projection of a hypercubic lattice in five and four dimensions, respectively \cite{deBruijn_1981,deBruijn_1981_2,Socolar_1989,Duneau_1989}. The remaining space after projection onto real space is called the perpendicular space. 
Each lattice site of a hypercube can be projected onto real space as well as the perpendicular space, and so a QC can be represented in terms of the projected lattice sites in the perpendicular space.
The three-dimensional perpendicular ($\tilde{x}\tilde{y}\tilde{z}$) space of Penrose QCs consists of four pentagonal cross sections of a rhombic icosahedron formed by the five projected basis vectors 
with four $\tilde{x}\tilde{y}$ planes equally spaced in the $\tilde{z}$ direction \cite{Jaric_1986}.
Those four pentagonal planes at $\tilde{z}=-n/\sqrt{5}$ ($n=1,2,3,4$) are illustrated in Fig.~\ref{fig:figPenAB}(c). 
The perpendicular space of AB QCs shown in Fig.~\ref{fig:figPenAB}(d) is two-dimensional, onto which all hypercubic lattice points are projected in the interior of an octagon associated with the four projected basis vectors.
Shown in Fig.~\ref{fig:figPenAB}(e) and (f) are the perpendicular space of Penrose QCs at $\tilde{z}=-3/\sqrt{5}$ and $-4/\sqrt{5}$, respectively.
The coordination number $Z_i$, which is the number of neighbouring sites linked to a given lattice site $i$ in a QC, varies from 3 to 7 (8) in Penrose (AB) QCs, and is plotted in Fig.~\ref{fig:figPenAB}(d-f).
The perpendicular space at $\tilde{z}=-2/\sqrt{5}$ ($-1/\sqrt{5}$) is identical including the distribution of $Z_i$ to that at $\tilde{z}=-3/\sqrt{5}$ ($-4/\sqrt{5}$) except that it is rotated by $\pi/5$ ($-\pi/5$).

In the perpendicular space, lattice sites of a QC with similar local environment are placed closely together; specifically, those with different coordination numbers are grouped in different sections in the perpendicular space. This can be seen in Fig.~\ref{fig:figPenAB}(d-f), where within each of the sections separated by black lines the coordination number $Z_i$ is uniform. One can plot a physical quantity of a QC in the perpendicular space and if the values of the physical quantity are distributed in the same or a similar manner as the coordination number, one can conclude that there is a direct correlation between the local environment of lattice sites and the physical quantity. Thus, the perpendicular-space representation of a QC is useful for discerning local vs. nonlocal effects on physical quantities. 
Furthermore, self-similarity of a QC can be studied easily in the perpendicular space. For example, by connecting only those lattice sites of a Penrose QC that are inside the inner pentagon in each of the four perpendicular space (\emph{not} the innermost one at $\tilde{z}=-2/\sqrt{5}$ and $-3/\sqrt{5}$), one obtains another Penrose QC with a larger link length. The sites of this larger QC form the same four patterns of the distribution of $Z_i$ in the perpendicular space as the original, now within each of the inner pentagons; except that the patterns at $\tilde{z}=-1/\sqrt{5}$ ($-2/\sqrt{5}$) and $\tilde{z}=-4/\sqrt{5}$ ($-3/\sqrt{5}$) are swapped.
Thus, the procedure can be repeated to obtain larger and larger Penrose QCs.

We use the tight-binding model for 2D TSC with broken time-reversal symmetry \cite{Sato_2009_3,Sato_2010} generalized for QCs \cite{Ghadimi_2021}, where the necessary ingredients are Rashba SO coupling, Zeeman field, and $s$-wave superconductivity. Our mean-field (BdG) Hamiltonian can be written as \cite{Goertzen_2017}
\begin{eqnarray}
\label{eq:Ham_generalized_KT}
{\mathcal H}&=&\sum_{i,j_i,\sigma}t_{i j_i}c^{\dagger}_{i\sigma}c_{j_i \sigma}
+\sum_{i,\sigma}\left( -\tilde{\mu} + \tilde{h}_{\sigma} + V^{\rm{(H)}}_{i\bar{\sigma}} - \bar{V}^{\rm{(H)}}_{\bar{\sigma}} \right)c^{\dagger}_{i\sigma}c_{i\sigma}
\nonumber\\
&+&\frac{\alpha}{2} \left( \sum_{i,j_{i}} e^{\imath \theta_{j_{i}}} c^{\dagger}_{j_{i}\downarrow}c_{i\uparrow} + \rm{H.c.} \right)\nonumber\\
&+&\sum_{i}\left( \Delta_{i}c^{\dagger}_{i\uparrow}c^{\dagger}_{i\downarrow} + \rm{H.c.} \right),
\end{eqnarray}
where $c^{\dagger}_{i\sigma}$ ($c_{i\sigma}$) creates (annihilates) an electron with spin $\sigma$ at site $i$, $t_{i j_i}$ is the probability amplitude for the electron to hop from site $j_i$ to site $i$, $\tilde{\mu}$ is the ``dressed'' chemical potential, and $\tilde{h}_\sigma=-\tilde{h}$ ($+\tilde{h}$) for $\sigma=\uparrow$ ($\sigma=\downarrow$) with the effective Zeeman energy $\tilde{h}$:
\begin{eqnarray}
&&\tilde{\mu} = \mu -\frac{\bar{V}^{\rm{(H)}}_\uparrow + \bar{V}^{\rm{(H)}}_\downarrow}{2}\,, \label{eq:effective_mu}\\
&&\tilde{h} = h + \frac{\bar{V}^{\rm{(H)}}_\uparrow - \bar{V}^{\rm{(H)}}_\downarrow}{2}\,. \label{eq:effective_h}
\end{eqnarray}
Here $\mu$ and $h$ are the ``bare'' chemical potential and Zeeman energy, respectively. 
We consider the electron hopping only along links connecting vertices from site $i$ to one of its ``nearest-neighbour'' sites $j_i$ with $t_{i j_i}\equiv -t$ and SO coupling constant $\alpha$. The angle between the $x$ axis and the vector connecting site $i$ to site $j_i$ is denoted as $\theta_{j_{i}}\in \left[-\pi,\pi\right)$ and $\imath=\sqrt{-1}$. 
Assuming intrinsic effective attraction among electrons with uniform coupling constant $U<0$, we calculate the superconducting order parameter and the Hartree potential created by electrons with spin $\sigma$ (and felt by those with spin $\bar{\sigma}\ne \sigma$) at site $i$ as
\begin{equation}\label{eq:meanfields}
\Delta_{i} = U \langle c_{i \downarrow} c_{i \uparrow} \rangle ,\quad\quad
V_{i \sigma}^{\rm{(H)}} = U \langle c_{i \sigma}^\dagger c_{i \sigma} \rangle 
\end{equation}
and the site average, $\bar{V}^{\rm{(H)}}_{\sigma} = (1/N)\sum_{i}V^{\rm{(H)}}_{i{\sigma}}$. For a given $h$ there are more electrons with spin up than down and $\bar{V}^{\rm{(H)}}_\uparrow < \bar{V}^{\rm{(H)}}_\downarrow$ in Eq.~(\ref{eq:effective_h}) and thus, the average Hartree potential effectively reduces the Zeeman field. 
We solve the BdG equations with the Hamiltonian in Eq.~(\ref{eq:Ham_generalized_KT}) for a QC with $N$ sites directly in real space and solve for both mean fields in Eq.~(\ref{eq:meanfields}) 
self-consistently, until they satisfy the convergence criteria,
\begin{equation}\label{eq:convergence}
\frac{\|\vec{\Delta}^{(l)}-\vec{\Delta}^{(l-1)}\|}{\|\vec{\Delta}^{(l-1)}\|} < \delta
\end{equation}
with $\delta=10^{-6}$--$10^{-8}$, where $\vec{\Delta}^{(l)}$ is a complex vector of length $N$ containing $\{\Delta_i\}$ at the $l$th iteration, and similarly in terms of real vectors of length $N$ for $\{V_{i \sigma}^{\rm{(H)}}\}$. The results shown below are for temperature $0.01t$, which are practically the same as for zero temperature.

Once the mean fields are converged, we determine the topological nature of the system by calculating the Bott index \cite{Loring_2010,Loring_2019}. The Bott index $B\in\mathbb{Z}$ is equivalent to the first Chern number for periodic systems in the thermodynamic limit \cite{Ge_2017,Yang_2019}. Calculation of the Bott index requires periodic boundary conditions (PBC), but not translational symmetry, and the Bott index can identify the bulk topology of disordered or aperiodic systems.  
The Bott index has been used for characterising topological states in QCs
\cite{Bandres_2016,Ghadimi_2021,Huang_2018,Huang_2018_2,Huang_2019}.
To calculate the Bott index, we first obtain all the eigenvectors for negative energy of
the BdG Hamiltonian in Eq.~(\ref{eq:Ham_generalized_KT}) with the converged order parameter and Hartree potential. 
We then construct the projection operator of the occupied states,
\begin{equation}
  \label{eq:occupation_projector}
  P = \sum_{\epsilon_{n}<0}\Ket{\psi_{n}}\Bra{\psi_{n}},
\end{equation}
and the projected position operators,
\begin{equation}
  \label{eq:projected_position_x}
  U_{X} = P e^{\imath 2 \pi X} P + (I-P), \quad U_{Y} = P e^{\imath 2 \pi Y} P + (I-P),
\end{equation}
where $I$ is the identity operator.
Here we define the position operators $X$ and $Y$ by mapping a 2D QC onto the surface of a torus:
\begin{equation}
  \label{eq:position_x}
  X = {\rm{Diag}}\left[x_{1},x_{1},\ldots,x_{N},x_{N},x_{1},x_{1},\ldots,x_{N},x_{N} \right],
\end{equation}
where $x_i\in\left[ 0, 1 \right)$ is the normalised $x$ coordinate of site $i$ and similarly for $Y$. 
The Bott index is given by
\begin{equation}
  \label{eq:Bott_index}
  B = \frac{1}{2\pi}\,{\rm{Im}} {\rm{Tr}} \ln{\left( U_{Y}U_{X}U^{\dagger}_{Y}U^{\dagger}_{X} \right)}.
\end{equation}
We perform our self-consistent calculations on Penrose and AB approximants, which PBC can be used for.

\begin{figure}[tbh]
\includegraphics[width=8cm]{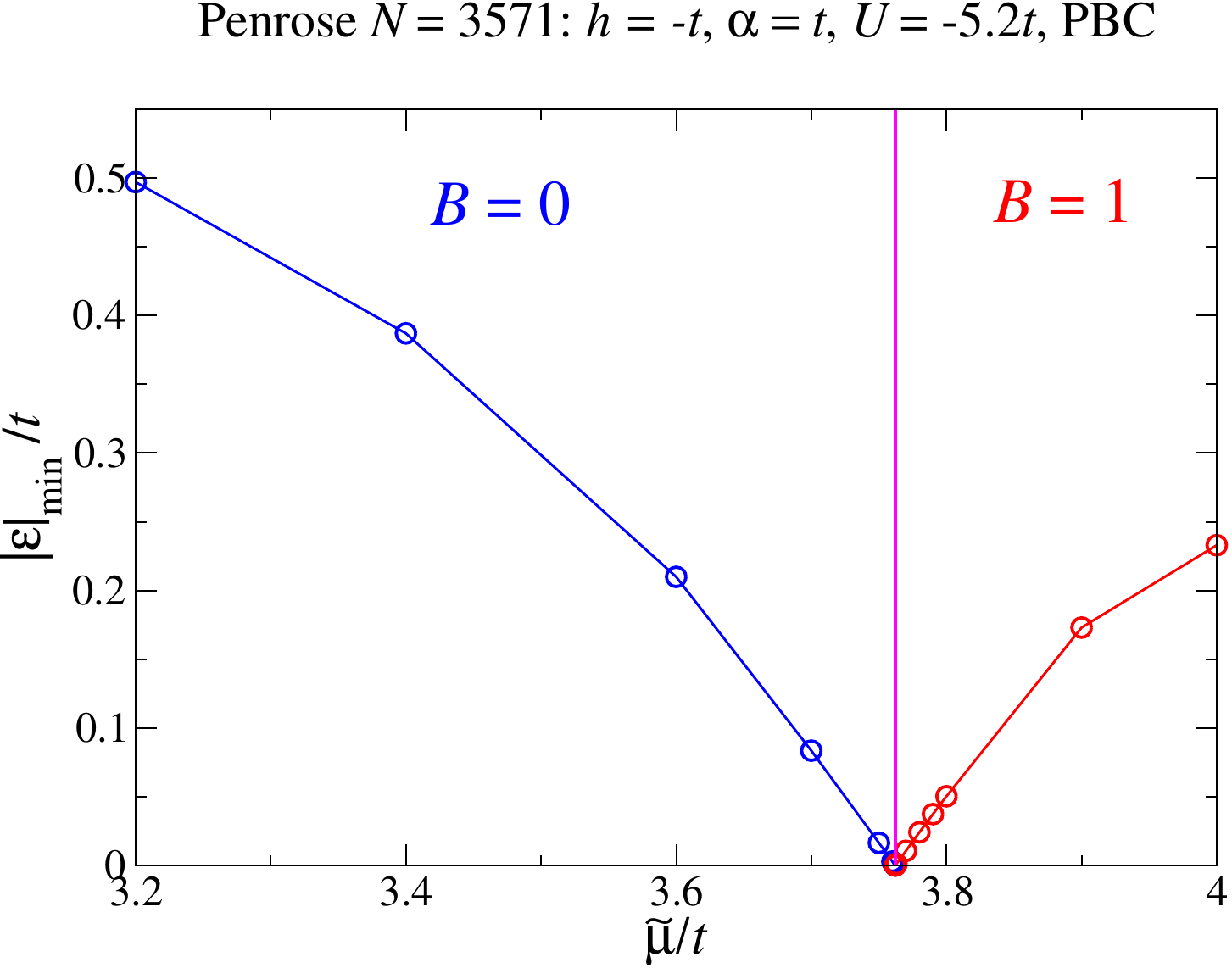}
\caption{Lowest absolute energy eigenvalue $|\epsilon|_{{\rm{min}}}/t$ as a function of the chemical potential $\tilde{\mu}/t$ for a Penrose QC with $N=3571$, $h=-t$, $\alpha=t$, $U=-5.2t$, and PBC.
\label{fig:PenBott_mutilde}}
\end{figure}

\begin{figure}[tbh]
\includegraphics[width=8cm]{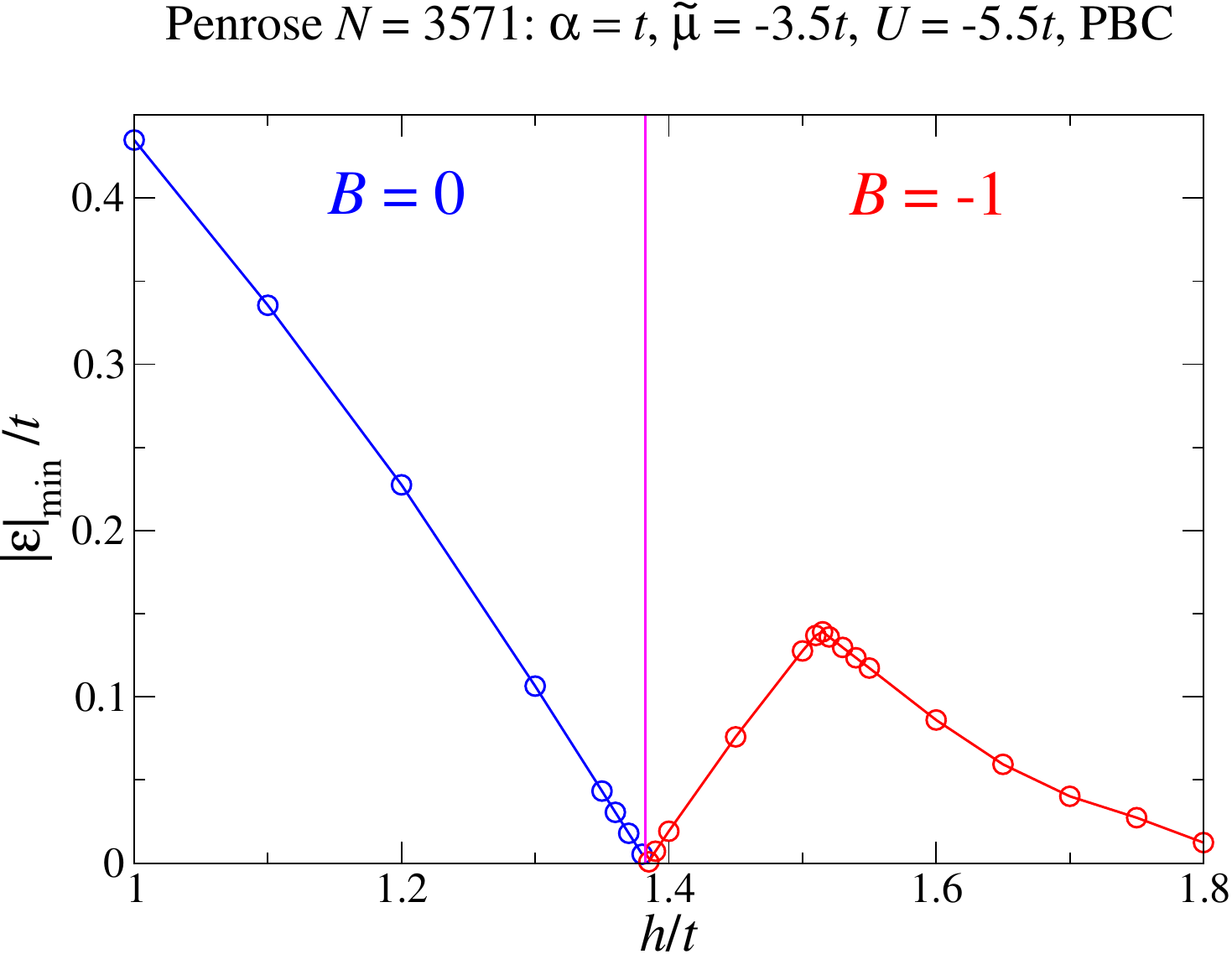}
\caption{Lowest absolute energy eigenvalue $|\epsilon|_{{\rm{min}}}/t$ as a function of the Zeeman energy $h/t$ for a Penrose QC with $N=3571$, $\alpha=t$, $\tilde{\mu} = -3.5t$, $U=-5.5t$, and PBC.
\label{fig:PenBott_h}}
\end{figure}

\section{\label{sec:results}RESULTS}
\subsection{\label{sec:PenQCs}Penrose quasicrystals}

\subsubsection{\label{sec:PenTopoPhase_Pen}Topological phase transitions}

In Ref.~\cite{Ghadimi_2021}, by assuming a constant superconducting order parameter $\Delta_i\equiv\Delta$, topological phase transitions were found to occur as a function of the chemical potential $\mu$ near the lower band edge for fixed $\alpha$ and $h$ in both Penrose and AB QCs. These phase transitions follow the same condition for closing of the bulk spectral gap, $(W+\mu)^2+\Delta^2 = h^2$, as for a square lattice \cite{Sato_2010}, where half of the kinetic-energy bandwidth $W$ is obtained numerically.
When it is solved for self-consistently, the order parameter $\Delta_i$ is not uniform and thus, there is no such gap-closing condition. Moreover, $\Delta_i$ acquires a non-negligible imaginary part at each lattice site as the inhomogeneity of the quasicrystalline lattice has similar effects as nonmagnetic impurities on the underlying $p\pm \imath p$ nature of TSC \cite{Goertzen_2017}.
Here we show in terms of self-consistent solutions the occurrence of a topological phase transition as $\tilde{\mu}$ is varied. Figure~\ref{fig:PenBott_mutilde} presents the lowest absolute value of the quasiparticle excitation energy $|\epsilon|_{{\rm{min}}}$ as a function of $\tilde{\mu}/t$ for a 3571-site Penrose QC for $h=-t$, $\alpha=t$, and $U=-5.2t$ with PBC. The spectral gap closes at $\tilde{\mu} \simeq 3.7625t$, where the Bott index changes from $B=0$ to $B=1$. 
For all values of $\tilde{\mu}$ studied here, the converged order parameter $\Delta_i$ is nonzero at all sites and the real part of the site average, $\bar{\Delta} = (1/N)\sum_i \Delta_i$, decreases monotonically from ${\rm Re}\bar{\Delta}\simeq 0.8t$ at $\tilde{\mu}=3.2t$ to ${\rm Re}\bar{\Delta}\simeq 0.3t$ at $\tilde{\mu}=4t$.
Comparing Fig.~\ref{fig:PenBott_mutilde} with Fig.~\ref{fig:sqBott_mutilde} for a square lattice with comparable $N$, it can be seen that $\tilde{\mu}$ at which the topological phase transition occurs is pushed up to a higher value in the Penrose QC. This makes sense from the fact that the kinetic-energy levels are more spread out in Penrose (as well as AB) QCs and go outside of the range $[-4t,4t]$ as for a square lattice \cite{Ghadimi_2021}.

By varying the Zeeman energy $h$ for fixed $\alpha$ and $\tilde{\mu}$, we also find a topological phase transition in our self-consistent solutions. This is illustrated for a 3571-site Penrose QC for $\alpha=t$, $\tilde{\mu} = -3.5t$, and $U=-5.5t$ with PBC in Fig.~\ref{fig:PenBott_h}, where the lowest absolute excitation energy $|\epsilon|_{{\rm{min}}}$ is plotted as a function of $h/t$. The spectral gap closes at $h\simeq 1.3845t$, where the Bott index changes from 0 to $-1$. As $h$ varies from $h=t$ to $h=1.8t$, the effective Zeeman energy $\tilde{h}$ increases from $\tilde{h}\simeq 0.86t$ to $\tilde{h}\simeq 1.35t$ and the real part of the site-averaged order parameter 
monotonically decreases from ${\rm Re}\bar{\Delta}\simeq 0.8t$ to ${\rm Re}\bar{\Delta}\simeq 0.1t$, while $\Delta_i$ remains nonzero at all sites.
The overall behaviour of the spectral gap as a function or $h$ or $\tilde{h}$ (not shown) around the phase transition and in the topological region with $B=-1$ is similar to that in a square lattice, as can be seen in Fig.~\ref{fig:sqBott_h}. However, compared to the square lattice, stronger Zeeman field is required for the transition to occur, and the spectral gap in the $B=-1$ region is significantly smaller in the QC and closes at $h\simeq 1.87t$.
%reduced to $5\times 10^{-4}t$ at $h=1.86t$. 
The TSC phase ceases to exist with vanishing order parameter ($|\Delta_i| \lesssim 10^{-5}t$) at all sites for $h\simeq 1.88t$ in the 3571-site Penrose QC.

\begin{figure}[tbh]
\includegraphics[width=8cm]{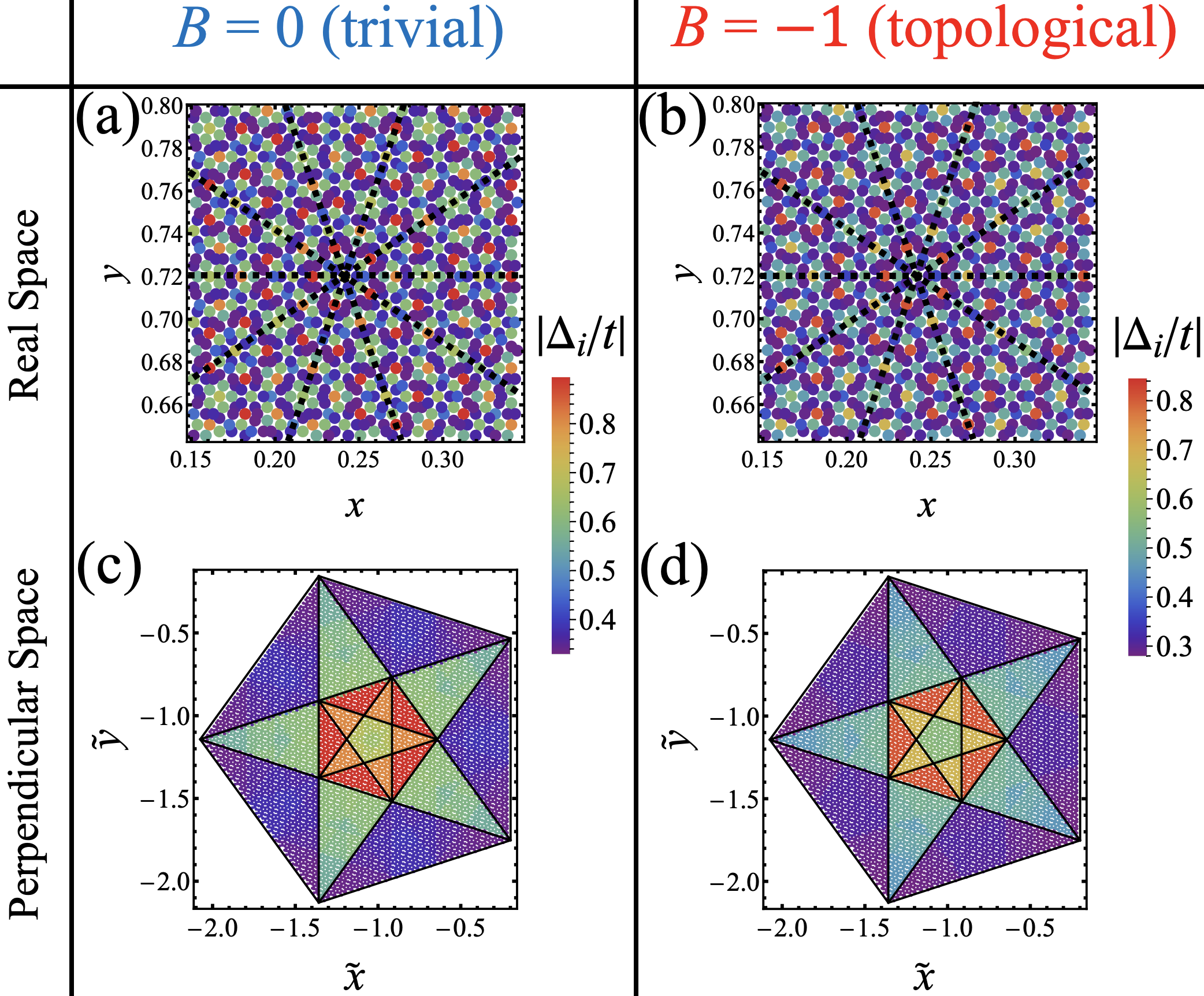}
\caption{$|\Delta_{i}/t|$ around a high-symmetry site in (a,b) real space and (c,d) the perpendicular space at $\tilde{z}=-3/\sqrt{5}$ for (a,c) $\tilde{\mu} = -3.4t$ ($B=0$) and (b,d) $\tilde{\mu} = -3.5t$ ($B=-1$) in a 24476-site Penrose QC with PBC, for $h = 1.5t$, $\alpha = t$, and $U = -5.5t$.
\label{fig:Pen24476AbsDeltaOriginal_rel}}
\end{figure}

\subsubsection{\label{sec:DistSCparam_Pen}Distribution of the superconducting order parameter}

\begin{figure*}[tbh]
\includegraphics[width=12cm]{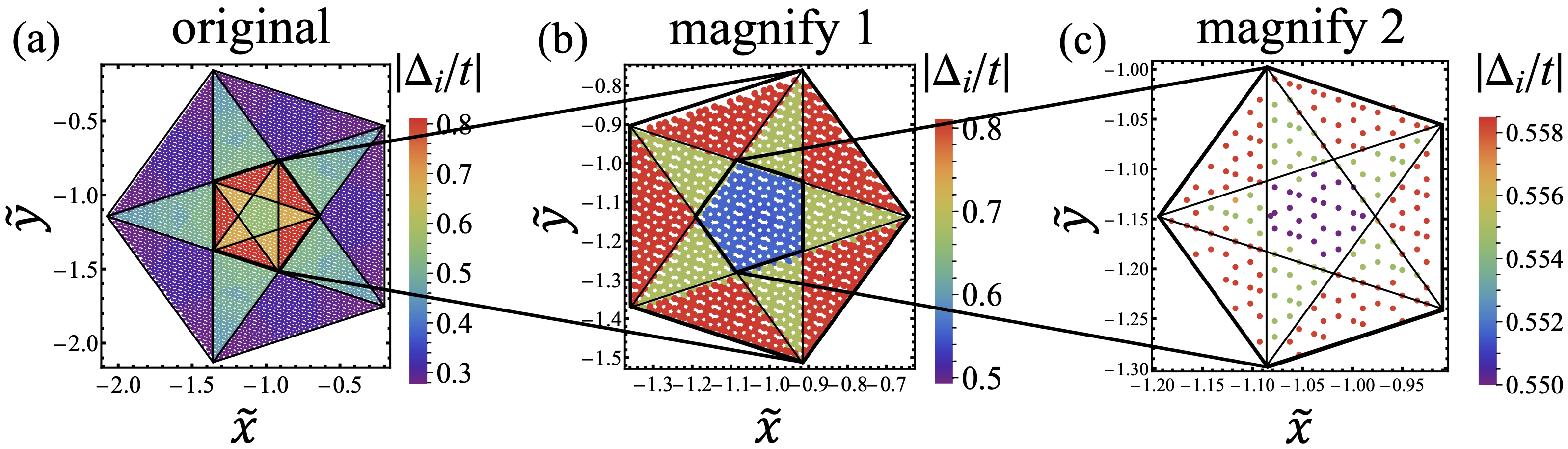}
\caption{(a) $|\Delta_{i}/t|$ in the perpendicular space at $\tilde{z}=-3/\sqrt{5}$ in Fig.~\ref{fig:Pen24476AbsDeltaOriginal_rel}(d). The inner pentagon marked by thick black lines in (a) is magnified in (b). The centre pentagon marked by thick black lines in (b) is magnified in (c).
\label{fig:Pen24476AbsDeltaMagnifyTwice_rel}}
\end{figure*}

When the BdG equations are solved self-consistently, the converged mean fields reflect the underlying quasicrystalline structure \cite{Sakai_2017,Sakai_2019,Araujo_2019,Sakai_2021} . In Fig.~\ref{fig:Pen24476AbsDeltaOriginal_rel} we show the magnitude of the converged order parameter in units of $t$ 
in a Penrose QC with $N=24476$ sites and PBC for $h = 1.5t$, $\alpha = t$, $U = -5.5t$, for (a,c) $\tilde{\mu} = -3.4t$ ($B=0$) and (b,d) $\tilde{\mu}=-3.5t$ ($B=-1$). In Fig.~\ref{fig:Pen24476AbsDeltaOriginal_rel}(a,b), $|\Delta_i|/t$ is plotted as a function of the normalised lattice coordinate $(x,y)$ in a small region around a high-symmetry lattice site, about which local fivefold rotational symmetry is indicated by black dashed lines. The Penrose QC also has local mirror symmetry about each black dashed line.
Other than slight differences in magnitude owing to the small difference in the chemical potential, 
there is no discernible difference in the distribution of $|\Delta_i|/t$ between $\tilde{\mu} = -3.4t$ and $\tilde{\mu}=-3.5t$: Yet, the former system is topologically trivial ($B=0$) and the latter nontrivial ($B=-1$). Clearly, the magnitude of the order parameter exhibits the same fivefold rotational symmetry and mirror symmetries as the underlying quasicrystalline lattice.

In Fig.~\ref{fig:Pen24476AbsDeltaOriginal_rel}(c,d) $|\Delta_i|/t$ for the entire Penrose QC is plotted as a function of $(\tilde{x},\tilde{y})$ in the perpendicular space at $\tilde{z}=-3/\sqrt{5}$. 
By comparing these plots and that of the coordination number in Fig.~\ref{fig:figPenAB}(e), 
it is clear that the magnitude of the order parameter is mostly determined by the coordination number, and the higher the coordination number, i.e., the more nearest-neighbour sites to hop to/from, the larger the $|\Delta_i|/t$.
This can be caused by a kinetic-energy gain and/or more stabilisation of TSC by spin-orbit coupling against the depairing effect of the Zeeman field, as a result of having more nearest-neighbour sites. For regular $s$-wave superconductivity, depending on the filling factor, the order parameter does not necessarily increase or linearly correlate with the coordination number \cite{Araujo_2019}; meaning that a kinetic-energy gain does not necessarily enhance $s$-wave superconductivity at a given site. However, for the particular case of $\tilde{\mu}=-3.5t$, $h=\alpha=0$, and $U=-4.95t$ (chosen so that the site average $\bar{\Delta}\approx 0.35t$ as for the topological case above) in a 3571-site Penrose QC, we also find larger $|\Delta_i|/t$ for higher coordination numbers.
In each of the sections that correspond to $Z_i=3$ in Fig.~\ref{fig:Pen24476AbsDeltaOriginal_rel}(c,d), slight increase in $|\Delta_i|/t$ around the line of symmetry of the outermost pentagon is visible. This may be due to some differences in the local environment of the three sites connected to each site $i$ with $Z_i=3$.
We note that the distribution of $|\Delta_i|/t$ in the perpendicular space still has the fivefold rotational symmetry and mirror symmetry about each line of symmetry of the pentagon as the distribution of $Z_i$ itself.

The magnitude of the order parameter also reflects the self-similar structure of Penrose QCs. This is illustrated in Fig.~\ref{fig:Pen24476AbsDeltaMagnifyTwice_rel} for the system presented in Fig.~\ref{fig:Pen24476AbsDeltaOriginal_rel}(b,d) for $B=-1$.
The perpendicular-space representation of $|\Delta_{i}/t|$ in Fig.~\ref{fig:Pen24476AbsDeltaOriginal_rel}(d) is shown again in Fig.~\ref{fig:Pen24476AbsDeltaMagnifyTwice_rel}(a). 
As explained in Sec.~\ref{sec:model}, the projected lattice sites inside the inner pentagon marked by thick black lines, along with those in the 
inner pentagons at the other three $\tilde{z}$ coordinates 
form a larger Penrose QC in real space. When the coordination numbers of this larger Penrose QC are plotted in the perpendicular space, the distribution of $Z_i$ in the inner pentagon in Fig.~\ref{fig:Pen24476AbsDeltaMagnifyTwice_rel}(a) will be the same as in Fig.~\ref{fig:figPenAB}(e) but rotated by $\pi/5$. In Fig.~\ref{fig:Pen24476AbsDeltaMagnifyTwice_rel}(b), the inner pentagon as well as its colour scale in (a) are magnified, where the distribution of $|\Delta_{i}/t|$ appears to be uniform within each $Z_i$ section. However, as the inner pentagon marked by thick black lines in (b) is further magnified along with its colour scale as shown in Fig.~\ref{fig:Pen24476AbsDeltaMagnifyTwice_rel}(c), 
it can be seen that roughly speaking, three different values of $|\Delta_i|/t$ are grouped in different sections separated by thin black lines. 
This distribution of roughly three values of $|\Delta_i|/t$ -- say, red, green, and purple in the descending order -- has the same pattern as the distribution of the coordination number of the larger Penrose QC, 
$Z_i=7$, 6, and 5 (see Fig.~\ref{fig:figPenAB}(e)), respectively, corresponding to the red, green and purple sections in Fig.\ref{fig:Pen24476AbsDeltaMagnifyTwice_rel}(c). Therefore, the magnitude of the order parameter has the same self-similar structure as the underlying Penrose QC.

\begin{figure}[tbh]
\includegraphics[width=8cm]{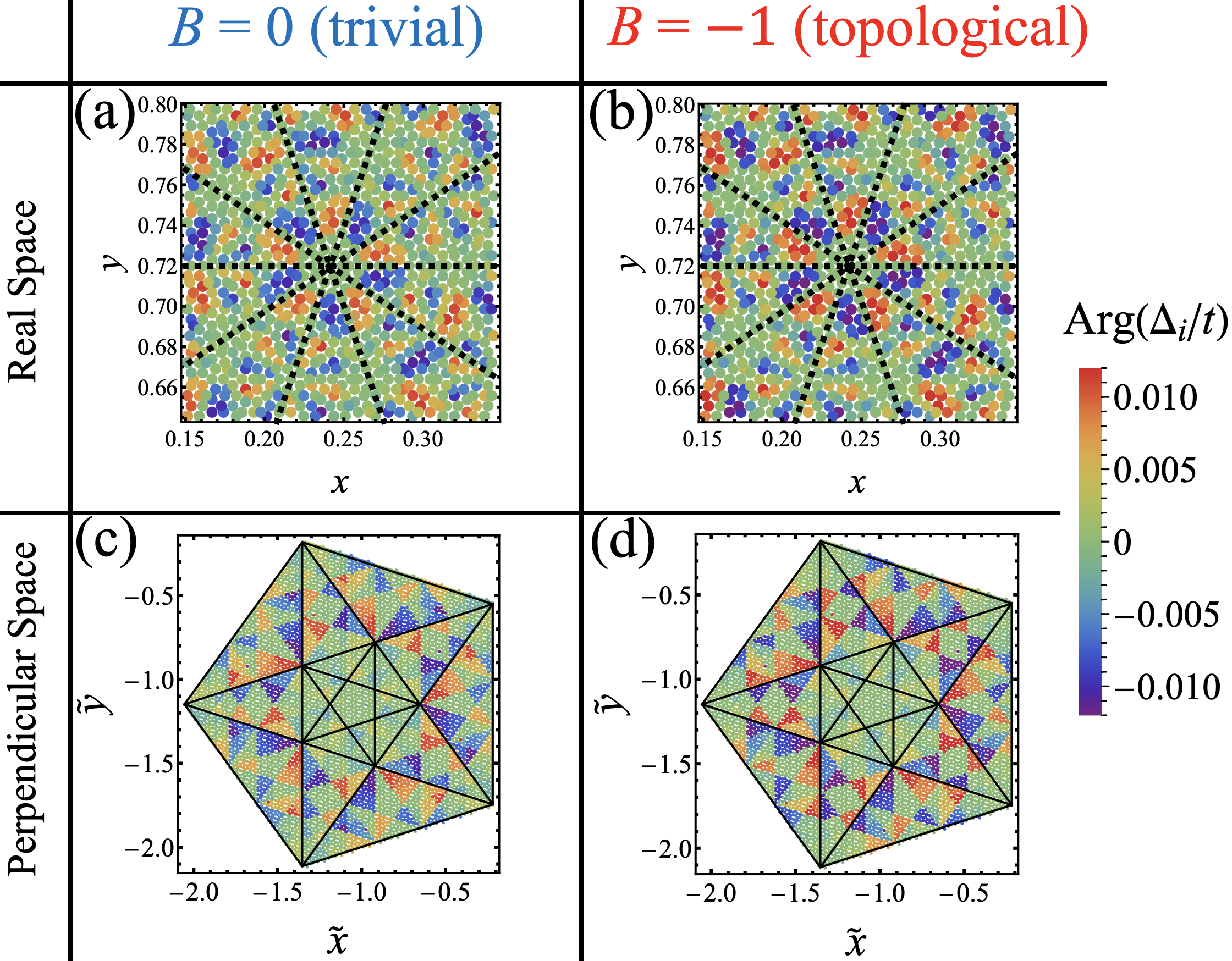}
\caption{${\rm{Arg}}\left(\Delta_{i}/t \right)$ around a high-symmetry site in real space (a,b) and in the perpendicular space at $\tilde{z}=-3/\sqrt{5}$ (c,d) for (a,c) $\tilde{\mu} = -3.4t$ ($B=0$) and $\tilde{\mu} = -3.5t$ ($B=-1$) in a 24476-site Penrose QC with PBC, for $h = 1.5t$, $\alpha = t$, and $U = -5.5t$.
\label{fig:Pen24476ArgDeltaOriginal_rel}}
\end{figure}

Analogously to Fig.~\ref{fig:Pen24476AbsDeltaOriginal_rel}, the phase of the converged order parameter, ${\rm{Arg}}\left(\Delta_{i}/t \right)$, for 
a 24476-site Penrose QC with PBC for $h = 1.5t$, $\alpha = t$, and $U = -5.5t$ is plotted as a function of $(x,y)$ around a high-symmetry site in Fig.\ref{fig:Pen24476ArgDeltaOriginal_rel} for (a) $\tilde{\mu} = -3.4t$ ($B=0$) and (b) $\tilde{\mu}=-3.5t$ ($B=-1$). Similarly to the magnitude distribution, there is no discernible difference in the phase distribution between the topologically trivial and non-trivial cases.
Once again, the black dashed lines indicate local fivefold rotational symmetry and are also lines of local mirror symmetry of the Penrose QC. The phase distribution in the perpendicular space at $\tilde{z}=-3/\sqrt{5}$ is shown for $\tilde{\mu} = -3.4t$ and $\tilde{\mu}=-3.5t$ in Fig.\ref{fig:Pen24476ArgDeltaOriginal_rel}(c) and (d), respectively. Comparing them with the distribution of the coordination number in Fig.~\ref{fig:figPenAB}(e), no apparent correlation between the phase of the order parameter and the coordination number can be seen. 
On the other hand, it can be seen in Fig.\ref{fig:Pen24476ArgDeltaOriginal_rel}(a,c) or (b,d) that ${\rm{Arg}}\left(\Delta_{i}/t \right)$ has the fivefold rotational symmetry of the QC both in real space and the perpendicular space. In contrast to $|\Delta_i|/t$, 
the phase has \emph{odd} mirror symmetry about each black dashed line in real space and about each line of symmetry of the pentagon in the perpendicular space. This makes sense as the phase must vary continuously as a function of spatial coordinate.

Finally, we note that the distribution of the converged Hartree potential, that is, the average electron density also reflects local fivefold rotational symmetry and mirror symmetries of the Penrose QC. It is also correlated with the coordination number, but in the manner opposite to the superconducting order parameter: The lower the coordination number $Z_i$, the higher the average electron density at site $i$ for each spin component. Namely, at lattice sites where the magnitude of the superconducting order parameter is minimum, the average electron density for both spin components is maximum, and vice versa. Thus, in a QC charge density modulations can coexist with TSC.

\subsubsection{\label{sec:Pen_Majorana}Majorana surface bound states}

\begin{figure}[tbh]
\includegraphics[width=8cm]{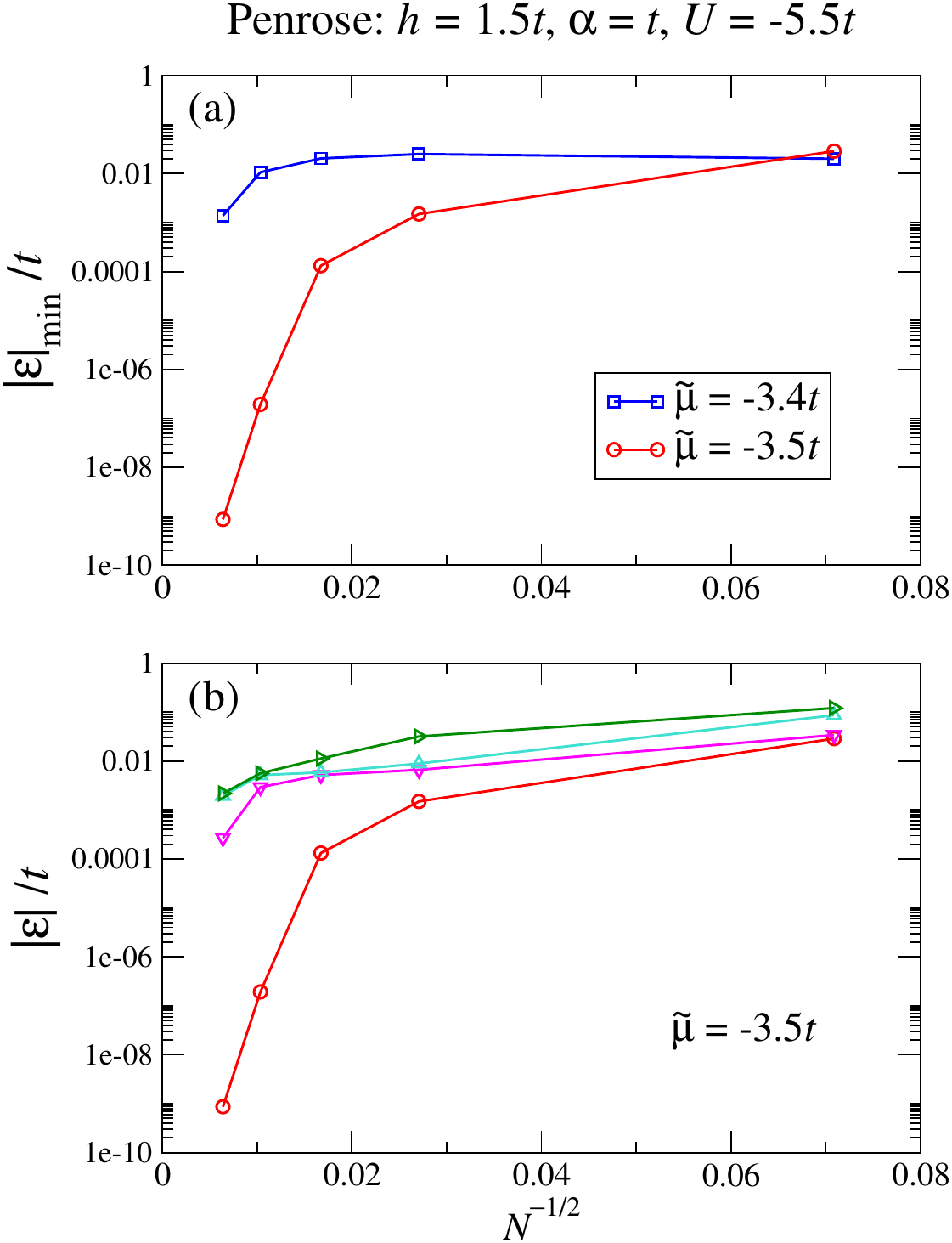}
\caption{(a) Absolute values of the minimum excitation energy $|\epsilon|_{\rm min}/t$ for $\tilde{\mu}=-3.4t$ ($B=0$) and $-3.5t$ ($B=-1$) and (b) the lowest four excitation energies $|\epsilon|/t$ for $\tilde{\mu}=-3.5t$ ($B=-1$) as a function of system size for a Penrose QC with OBC (PBC) in the $x$ ($y$) direction for $h=1.5t$, $\alpha=t$, and $U=-5.5t$. The number of sites are $N=199$, 1364, 3571, 9349, and 24476.\label{fig:Cylinder_PenAB_Majorana1}}
\end{figure}

When $B=\pm 1$, introducing two edges by using open boundary conditions (OBC) in the $x$ direction and PBC in the $y$ direction results in the appearance of a zero-energy Majorana bound state along each surface boundary. This is consistent with the bulk-edge correspondence, even though 
an index theorem for systems with translational symmetry, where a low-energy Hamiltonian can be approximated to be Dirac-like \cite{Fukui_2012}, 
may not be directly applicable to QCs.
In Fig.~\ref{fig:Cylinder_PenAB_Majorana1} we present lowest excitation energies as a function of system size in a Penrose QC with OBC and PBC in the $x$ and $y$ directions, respectively, for $h=1.5t$, $\alpha=t$, and $U=-5.5t$. A logarithmic scale is used for energy and the number of lattice sites are $N=199$, 1364, 3571, 9349, and 24476. In (a) we compare the lowest absolute energy $|\epsilon|_{\rm min}/t$ for $\tilde{\mu}=-3.4t$ and $-3.5t$, for which the Bott index $B=0$ and $-1$, respectively. Clearly, the minimum excitation energy approaches zero as the system size increases, $N^{-1/2}\rightarrow 0$, when $B=-1$. In contrast, there is no such trend when $B=0$. In Fig.~\ref{fig:Cylinder_PenAB_Majorana1}(b) the first four excitation energies are plotted as a function of $N^{-1/2}$ for $\tilde{\mu}=-3.5t$ ($B=-1$). It can be seen that only the lowest energy approaches zero as $N\rightarrow\infty$ and thus, this state is distinct from other quasiparticle states.
With particle-hole symmetry inherent in the BdG equations and the presence of two edges, when $B=\pm 1$, there are two such zero-energy states with energy $\pm |\epsilon|_{\rm min}$ which become degenerate in the thermodynamic limit.

\begin{figure}[tbh]
\includegraphics[width=8cm]{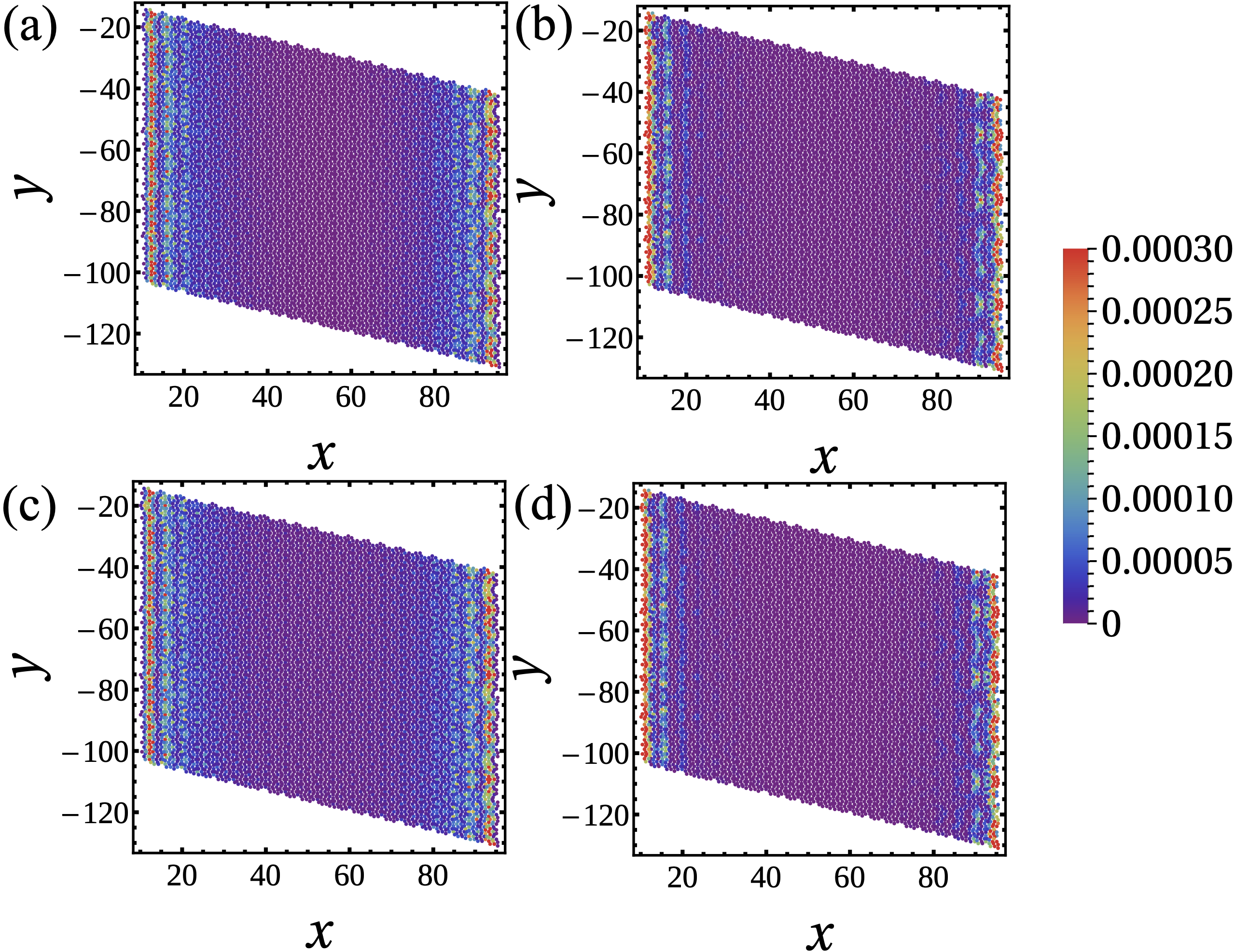}
\caption{(a,b) Electron and (c,d) hole amplitudes for spin (a,c) up and (b,d) down for the quasiparticle excitation with energy $\epsilon\simeq 10^{-4}t$ in a 9349-site Penrose QC with OBC and PBC, respectively, in the $x$ and $y$ directions, for $h=t$, $\alpha=t$, $U = -5.2t$, and  $\tilde{\mu} = 3.8t$.
\label{fig:Cylinder_PenAB_Majorana2}}
\end{figure}

Shown in Fig.~\ref{fig:Cylinder_PenAB_Majorana2} are the (a,b) electron and (c,d) hole amplitudes (magnitude squared) of the lowest-energy excitation for spin (a,c) up and (b,d) down in a Penrose QC with $N=9349$ and OBC (PBC) in the $x$ ($y$) direction for $h=t$, $\alpha=t$, $U = -5.2t$, and  $\tilde{\mu} = 3.8t$. The Bott index $B=-1$ in this system. This quasiparticle state, with energy $\epsilon\simeq 5\times 10^{-4}t$ ($\epsilon\simeq 2\times 10^{-5}t$ for $N=24476$), has the highest probability amplitudes along the two edges and is thus a linear superposition of two zero-energy surface bound states. We have chosen this system for presenting the probability amplitudes since it has a small spectral gap of about $0.05t$ (see Fig.~\ref{fig:PenBott_mutilde}) and hence a relatively long effective coherence length, so that the gradual overall decay of the amplitudes as one moves away from each edge is visible in Fig.~\ref{fig:Cylinder_PenAB_Majorana2}. It is also clear by comparing the plots in (a) and (c), or (b) and (d), that the electron and hole amplitudes are equal in magnitude for each spin component. 
Therefore, we have a zero-energy Majorana bound state along each edge.
And the larger the system and/or the shorter the effective coherence length, the less overlap in the middle of the system, if any, of the surface bound states along the two edges and $\epsilon\rightarrow 0$. In this limit, 
the two degenerate states can be superposed such that one state is a Majorana zero mode along one edge and the other state is another Majorana zero mode along the other edge.

\subsection{\label{sec:ABQCs}Ammann-Beenker quasicrystals}

\subsubsection{\label{sec:DistSCparam_AB}Distribution of the superconducting order parameter}

\begin{figure}[tbh]
\includegraphics[width=8cm]{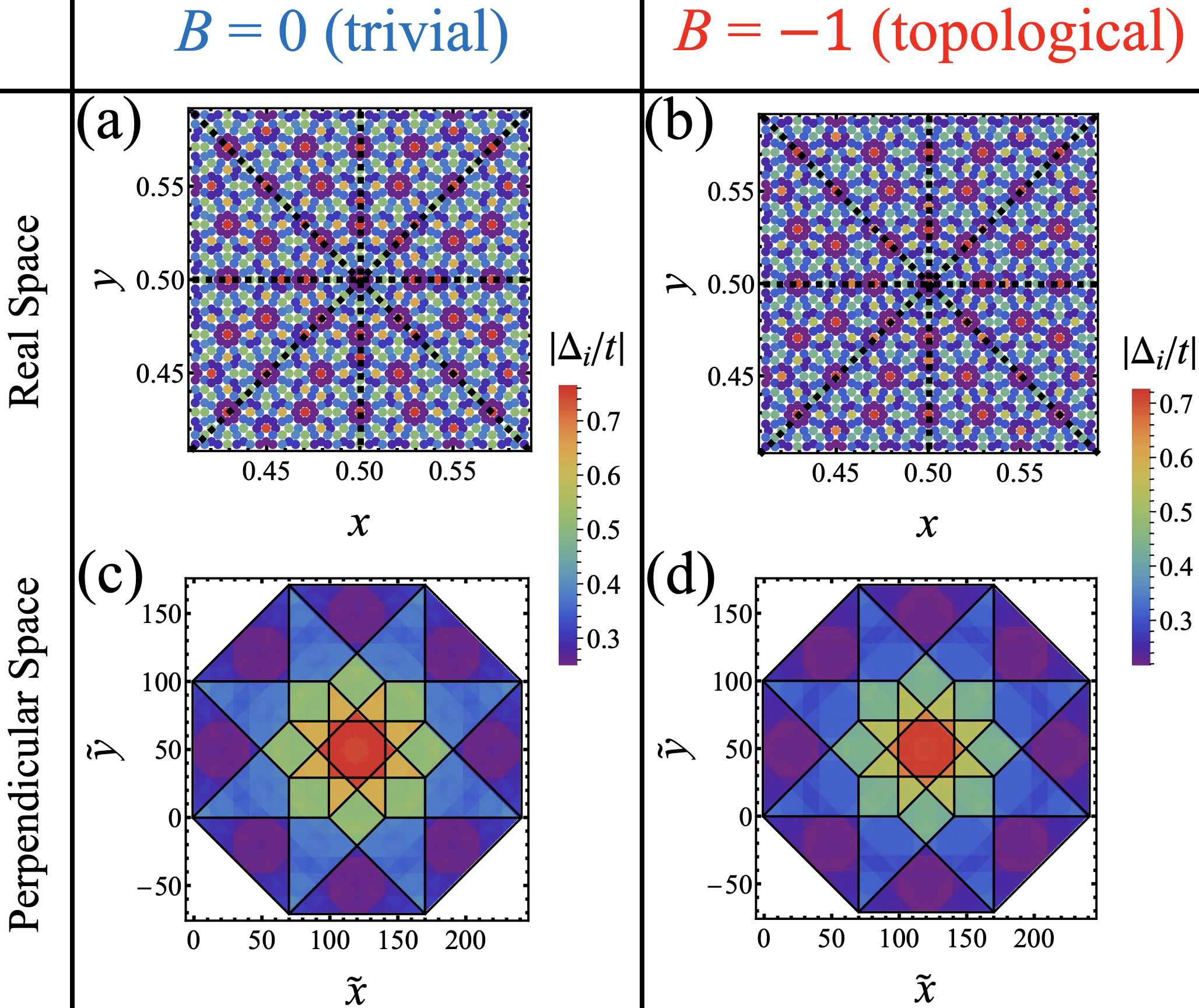}
\caption{$|\Delta_{i}/t|$ around a high-symmetry site in (a,b) real space and (c,d) the perpendicular space at $\tilde{z}=-3/\sqrt{5}$ for (a,c) $\tilde{\mu} = -3.4t$ ($B=0$) and (b,d) $\tilde{\mu} = -3.5t$ ($B=-1$) in a 47321-site AB QC with PBC, for $h = 1.5t$, $\alpha = t$, and $U = -5.5t$.
\label{fig:AB47321AbsDeltaOriginal_Jun2}}
\end{figure}

We have also solved the BdG equations self-consistently with the Hamiltonian in Eq.~(\ref{eq:Ham_generalized_KT}) on AB QCs. In Fig.~\ref{fig:AB47321AbsDeltaOriginal_Jun2}, the converged $|\Delta_{i}/t|$ is plotted in (a,b) real space as a function of the normalised coordinate $(x,y)$ around a high-symmetry site and (c,d) the perpendicular space at $\tilde{z}=-3/\sqrt{5}$, for an AB QC with $N=47321$ and PBC, for $h = 1.5t$, $\alpha = t$, and $U = -5.5t$; for (a,c) $\tilde{\mu} = -3.4t$ ($B=0$) and (b,d) $\tilde{\mu}=-3.5t$ ($B=-1$). Although they lead to topologically distinct states, 
there is no substantial difference in the distribution of the mean fields for the two values of $\tilde{\mu}$. 
We use inflation mapping to produce AB approximants \cite{Duneau_1989}. In an AB QC thus generated, the centre site has eightfold rotational symmetry locally around it, as indicated by black dashed lines in Fig.~\ref{fig:AB47321AbsDeltaOriginal_Jun2}(a) and (b).
The AB QC also possesses local mirror symmetry about each of these black dashed lines. It can be seen clearly that $|\Delta_{i}/t|$ exhibits the same rotational and mirror symmetries as the QC. Also in AB QCs, we find that the higher the coordination number $Z_i$, the larger the superconducting order parameter, $|\Delta_{i}/t|$. This can be seen by comparing Fig.~\ref{fig:figPenAB}(f) and Fig.~\ref{fig:AB47321AbsDeltaOriginal_Jun2}(c,d). In the latter figures, some patterns of slight variation in $|\Delta_{i}/t|$ within each section are visible, especially for $Z_i=3$ and 4. Still, eightfold rotational symmetry of the perpendicular space and mirror symmetry about each line of symmetry of the octagon are preserved. 

Similarly as we have seen in Penrose QCs, the phase of the order parameter, ${\rm Arg}\left(\Delta_i/t\right)$, 
which is uncorrelated with coordination number $Z_i$, 
reflects the eightfold rotational symmetry of AB QCs in real space as well as in the perpendicular space. On the other hand, it presents \emph{odd} mirror symmetry about each line of mirror symmetry of AB QCs both in real and perpendicular spaces. 
The average electron density for each spin component is also distributed according to the underlying structure of the AB QC, reflecting its local eightfold rotational and mirror symmetries; and the smaller the $Z_i$, the larger the average electron density for both spin components.

\subsubsection{\label{sec:AB_Majorana}Majorana vortex bound state}

\begin{figure}[tbh]
\includegraphics[width=8cm]{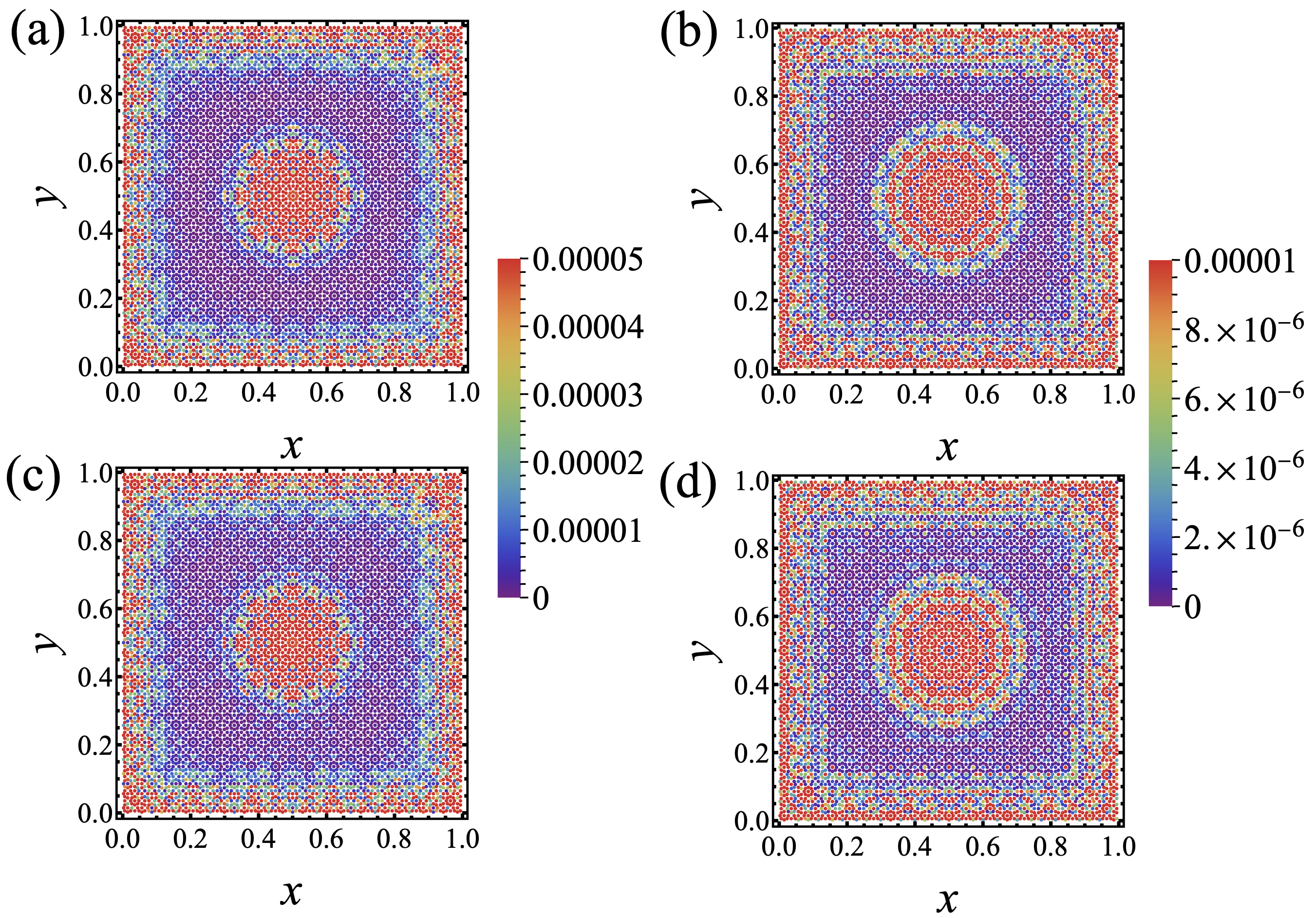}
\caption{(a,b) Electron and (c,d) hole amplitudes for spin (a,c) up and (b,d) down for the quasiparticle excitation with energy $\epsilon\simeq 4\times 10^{-5}t$ in a 8119-site AB QC with a vortex at the centre and OBC all around, for $h=1.5t$, $\alpha=t$, $U = -5.5t$, and $\tilde{\mu} = -3.5t$.
\label{fig:AB8119_vortex_KTsystems_vortexModes}}
\end{figure}

By placing a vortex at the centre of the system and using OBC all around, we find two zero-energy Majorana bound states, one at the vortex and the other along the edges, when $B=\pm 1$. We have solved for the mean fields self-consistently by introducing a phase winding of $2\pi$ in the superconducting order parameter about the centre site of AB QCs. 
Figure~\ref{fig:AB8119_vortex_KTsystems_vortexModes} shows the (a,b) electron and (c,d) hole amplitudes (magnitude squared) of the lowest-energy excitation with energy $\epsilon\simeq 4\times 10^{-5}t$ for spin (a,c) up and (b,d) down in an AB QC with $N=8119$ and OBC, for $h = 1.5t$, $\alpha = t$, $U = -5.5t$, $\tilde{\mu} = -3.5t$. The Bott index $B=-1$, as for the system presented in Fig.~\ref{fig:AB47321AbsDeltaOriginal_Jun2}(b,d).
It can be seen in all of Fig.~\ref{fig:AB8119_vortex_KTsystems_vortexModes} that the amplitudes are the highest around the vortex centre and along the edges, and the electron and hole amplitudes are (almost) the same at all sites for both spin up (a,c) and down (b,d). We have confirmed that the lowest-energy excitation with $\pm\epsilon$ is distinct from all higher-energy excitation and that $\epsilon$ approaches zero as $N$ increases. For the system presented in Fig.~\ref{fig:AB8119_vortex_KTsystems_vortexModes} but for $N=47321$, $\epsilon\simeq 5\times 10^{-8}t$. Moreover, as the system size increases and hence any overlap between the vortex and edge bound states vanishes, the electron and hole amplitudes (magnitude) become equal. Thus, in the thermodynamic limit, the two degenerate states with energy $\pm\epsilon\rightarrow 0$ can be superposed such that one Majorana zero mode is bound at the vortex and another Majorana zero mode is bound along the edges \cite{Bjoernson_2013}. The appearance of one Majorana zero mode per vortex when $B=\pm 1$ is consistent with the bulk-defect correspondence for periodic systems \cite{Teo_2010}, despite the lack of periodicity in QCs.

\subsection{\label{sec:impurities}Effects of nonmagnetic impurities}

Due to its $p\pm \imath p$ nature, TSC tends to be affected readily by nonmagnetic impurities \cite{impurities}, which can be incorporated as the single-particle potential $v_i=V_{\rm imp}$ at a given lattice site $i$, namely, %, that is, 
a local shift of the chemical potential by $-V_{\rm imp}$.
For a square lattice, it has been shown that a single nonmagnetic impurity can cause spin-polarised midgap excitation and even phase transitions of the ground state as the impurity potential $V_{\rm imp}$ is varied \cite{Goertzen_2017}. 
Furthermore, in terms of the self-consistently solved, average impurity self-energy, the work of Ref.~\cite{Nagai_2014} shows that as the concentration of nonmagnetic impurities increases, the spectral gap and the superconducting transition temperature are reduced. 
And the stronger the Zeeman field, the more drastic these effects of nonmagnetic impurities are on TSC \cite{Goertzen_2017,Nagai_2014}.

\begin{figure}[tbh]
  \includegraphics[width=9cm]{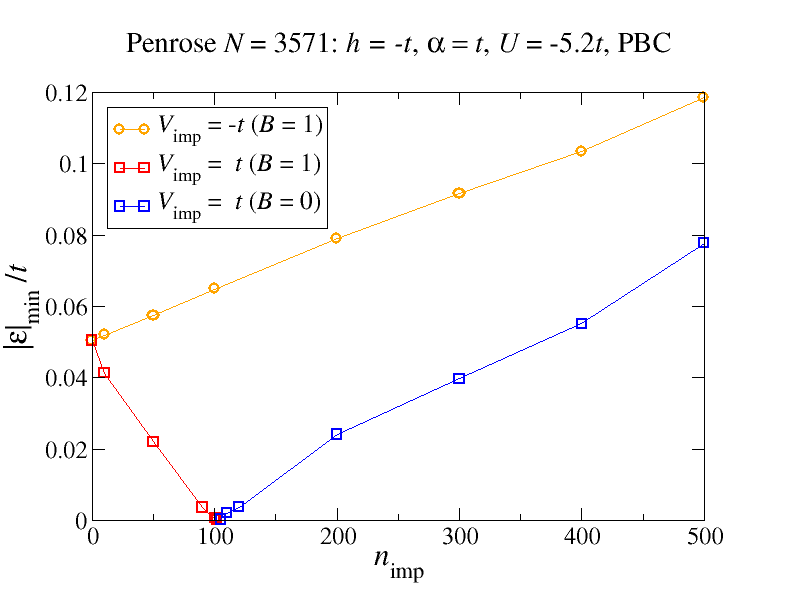}
\caption{Lowest absolute energy eigenvalue $|\epsilon|_{{\rm{min}}}/t$ as a function of the number $n_{\rm imp}$ of randomly distributed nonmagnetic impurities with $V_{\rm imp}=-t$ (orange circles) or $V_{\rm imp}=t$ (red and blue squares) in a Penrose QC with $N=3571$, $h=-t$, $\alpha=t$, $\tilde{\mu}=3.8t$, $U=-5.2t$, and PBC.
\label{fig:PT_imp}}
\end{figure}

What happens to TSC in QCs if nonmagnetic impurities are added randomly to the quasicrystalline lattice? We have examined the effects of nonmagnetic impurities with a constant potential $V_{\rm imp}$ distributed at $n_{\rm imp}$ random sites in both a square lattice and QCs. 
In Fig.~\ref{fig:PT_imp} we present the lowest absolute value of the quasiparticle excitation energy $|\epsilon|_{{\rm{min}}}$ as a function of $n_{\rm imp}$ with $V_{\rm imp}=-t$ (orange circles) and $V_{\rm imp}=t$ (red and blue squares) in a $3571$-site Penrose QC with PBC for $h=-t$, $\alpha=t$, $\tilde{\mu}=3.8t$, and $U=-5.2t$. For both $V_{\rm imp}=\pm t$, the same set of random sites $\{i\}$ was used to place the nonzero $v_i$ for a given $n_{\rm imp}$. For the repulsive potential $V_{\rm imp}=t$, the spectral gap monotonically decreases as $n_{\rm imp}$ increases, showing that TSC is suppressed by nonmagnetic impurities as expected. At $n_{\rm imp}=103$ the gap closes and the Bott index changes from 1 to 0; and as $n_{\rm imp}$ is increased further, the spectral gap increases in the trivial superconducting phase with $B=0$.

For the attractive potential $V_{\rm imp}=-t$, we have found the opposite effect that the spectral gap is enhanced as the impurities are added randomly to the QC, as can be seen in Fig.~\ref{fig:PT_imp}. The Bott index remains to be 1 and thus, the TSC phase is made more robust by randomly distributed nonmagnetic impurities. This may be reminiscent of a topological Anderson insulator \cite{Li_2009} as a result of a topological phase transition from a normal metal caused by disorder (typically a random shift of the chemical potential at each lattice site), with increasing energy gap as the disorder is made stronger.
In the current case, however, the increasing energy gap is basically due to the effective chemical potential of the system being increased as $n_{\rm imp}$ increases (see Fig.~\ref{fig:PenBott_mutilde}).
In fact, as $n_{\rm imp}$ is increased further, the spectral gap reaches its maximum value $\simeq 0.18t$ at $n_{\rm imp}\approx 1200$, beyond which it decreases gradually and is about $0.08t$ at $n_{\rm imp}=3,000$.
This is also consistent with the effective chemical potential of the system increasing,
as such dependence of the gap on $\tilde{\mu}$ is expected for $\tilde{\mu}>4t$ by comparing Fig.~\ref{fig:PenBott_mutilde} with Fig.~\ref{fig:sqBott_mutilde}.  

\begin{figure}[tbh]
\includegraphics[width=8cm]{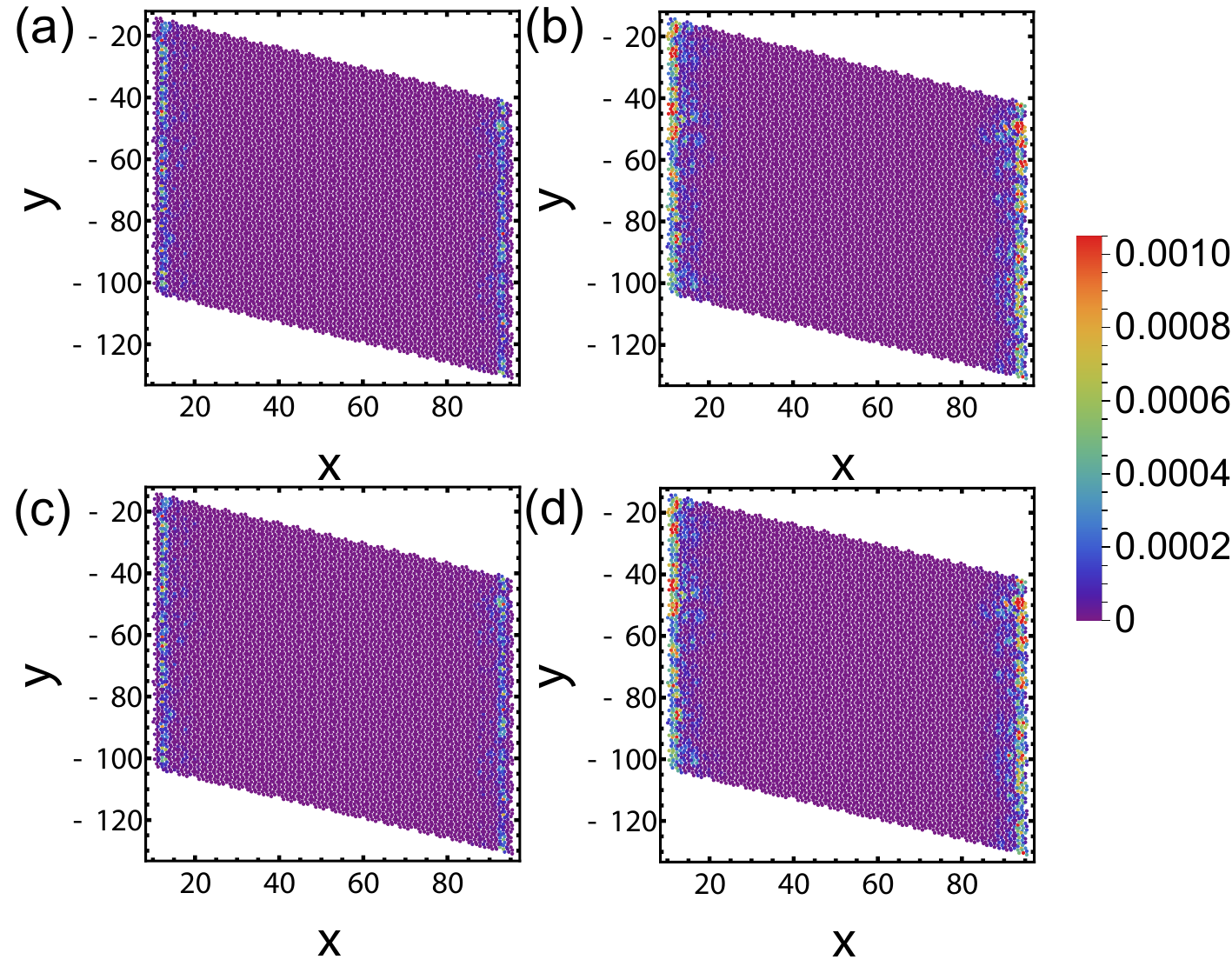}
\caption{(a,b) Electron and (c,d) hole amplitudes for spin (a,c) up and (b,d) down for the quasiparticle excitation with energy $\epsilon\simeq 10^{-10}t$ in the same system as in Fig.~\ref{fig:Cylinder_PenAB_Majorana2}, but with $4,000$ randomly distributed nonmagnetic impurities with $V_{\rm imp}=-t$.
\label{fig:Pen_Majorana_imp}}
\end{figure}

A larger spectral gap implies a shorter effective coherence length. This is demonstrated in Fig.~\ref{fig:Pen_Majorana_imp} for the same system presented in Fig.~\ref{fig:Cylinder_PenAB_Majorana2}, a 9349-site Penrose QC with OBC and PBC in the $x$ and $y$ directions, respectively, for $h=t$, $\alpha=t$, $U = -5.2t$, and  $\tilde{\mu} = 3.8t$, but with $n_{\rm imp}=4,000$ and $V_{\rm imp}=-t$. The energy of the Majorana edge modes is $\epsilon\simeq 5\times 10^{-4}t$ and $3\times 10^{-10}t$, respectively, without and with the nonmagnetic impurities. Comparing Figs.~\ref{fig:Cylinder_PenAB_Majorana2} and \ref{fig:Pen_Majorana_imp}, it is clear that the Majorana wave functions are confined much more strongly along and near each edge, and therefore, the two Majorana bound states at the two edges have negligible overlap.

We have also performed calculation for $h=1.5t$, $\alpha=t$, $\tilde{\mu}=-3.5t$, and $U=-5.5t$
in a $3571$-site Penrose QC with PBC. Compared with the system discussed above, the spectral gap in this system with no impurities is $0.127t$, which is much larger than $0.05t$ at $n_{\rm imp}=0$ in Fig.~\ref{fig:PT_imp} (see also Figs.~\ref{fig:PenBott_mutilde} and \ref{fig:PenBott_h}), and the signs of $h$ and $\tilde{\mu}$ are opposite. When the sign of the chemical potential is changed, the effects of the opposite signs of $V_{\rm imp}$ are interchanged \cite{Goertzen_2017}, so that the attractive potential only has the detrimental effect. For $V_{\rm imp}=-t$, because of the larger gap in the current system, the TSC phase persists up to $n_{\rm imp}=431$ before turning into the trivial superconducting phase. At the same time, as $|h|/t$ is larger, the effects of the nonmagnetic impurities are more drastic in that at $n_{\rm imp}=100$ the spectral gap is already reduced to $0.016t$.
Likewise, for $V_{\rm imp}=t$, the spectral gap increases only slightly up to the maximum value $\simeq 0.135t$ at $n_{\rm imp}=145$ and then decreases though gradually to $0.055t$ at $n_{\rm imp}=3,000$. When the Zeeman energy is increased to $h=1.7t$, the spectral gap ($0.04t$ at $n_{\rm imp}=0$) 
starts to decrease monotonically with increasing $n_{\rm imp}\gtrsim 10$. 
Therefore, the enhancement of TSC by randomly distributed nonmagnetic impurities can occur
only when the Zeeman field is not too strong. 

Finally, we have performed analogous calculation on a $60\times 60$ square lattice with PBC for the two parameter sets used above, 
with $V_{\rm imp}=t$ or $-t$. In contrast to the results discussed above,
for both parameter sets and for either sign of $V_{\rm imp}$, we have only found the detrimental effect on TSC by nonmagnetic impurities placed at random sites in a square lattice. The spectral gap starts to decrease as soon as $n_{\rm imp}$ is nonzero and continues to decrease as $n_{\rm imp}$ increases, though more slowly for $V_{\rm imp}=-t$ and $V_{\rm imp}=t$, respectively, for the former and latter parameter sets.
For the former parameter set, the reduction of the gap can be understood again as due to a shift in the effective chemical potential of the system by looking at Fig.~\ref{fig:sqBott_mutilde}: Without any impurity, the spectral gap is maximum around $\tilde{\mu}=3.8t$ and shifting $\tilde{\mu}$ in either direction would decrease the gap. For the latter parameter set, even though the spectral gap in the clean system is twice as large as that in the 3571-site Penrose QC for $h=1.5t$ (see Fig.~\ref{fig:sqBott_h}), the relatively large Zeeman energy presumably dominates the effects of the nonmagnetic impurities.

\section{\label{sec:conclusion}CONCLUSION}

In summary, we have shown the stable occurrence of 2D TSC with broken time-reversal symmetry \cite{Sato_2009_3,Sato_2010} in Penrose and AB QCs. Assuming an effective attractive interaction among electrons that leads to $s$-wave superconductivity, intrinsic Rashba spin-orbit coupling, and Zeeman field, we have solved the BdG equations self-consistently for the superconducting order parameter and the spin-dependent Hartree potential. The Bott index $B$ signifies the topological nature of the quasicrystalline system. By varying the chemical potential or Zeeman field, we have observed topological phase transitions where $B$ changes between 0 and 1 or $-1$, just as in a square lattice where the bulk spectral gap closes at one of the high-symmetry points in the Brillouin zone \cite{Sato_2010}.
We have demonstrated the appearance of a Majorana zero mode per surface boundary or vortex when $B=\pm 1$. This is consistent with the bulk-boundary \cite{Fukui_2012} and bulk-defect \cite{Teo_2010} correspondence, respectively, for periodic systems.

Similarly as found in other tight-binding models in Penrose \cite{Sakai_2017,Sakai_2019} and AB \cite{Araujo_2019,Sakai_2021} QCs, the self-consistently obtained mean fields are inhomogeneous, reflecting the geometry of the QC and the local environment of each lattice site, whether $B$ is zero or nonzero.
In Penrose (AB) QCs, both magnitude and phase of the superconducting order parameter and each spin component of the Hartree potential exhibit fivefold (eightfold) rotational symmetry in real space, locally about a high-symmetry point, and in the perpendicular space. The magnitude of the order parameter and the Hartree potential also present mirror symmetries of the underlying QC in both real space and the perpendicular space. The phase of the order parameter varies continuously across each line of mirror symmetry in the QC and possesses odd, in place of even, mirror symmetries. 
Furthermore, we have shown that the magnitude of the converged order parameter inherits the self-similarity of the QC. It is expected that the Hartree potential, i.e., the average electron density for each spin component has the self-similar distribution as well, as has been shown for the extended repulsive Hubbard model on AB QCs in the Hartree-Fock approximation \cite{Sakai_2021}.

With the order parameter and the average electron density varying from site to site, we have found superconductivity coexisting with charge density modulations in both Penrose and AB QCs. For the chemical potential close to the top or bottom of the kinetic-energy band, the higher the coordination number, the larger the magnitude of the order parameter and the smaller the average electron density for each spin component. 

Throughout this work, we have used the vertex model where the electron hops from site to site along the edge of a tile constituting the quasiperiodic tiling and as all edges are of the same length, this makes both Penrose and AB QCs bipartite. This may not be realistic in modelling real systems in that the tile edge is not necessarily the shortest link that connects two lattice sites. However, in the work of Ref.~\cite{Ghadimi_2021} it has been shown for a uniform order parameter that adding shortest-distance links, thus breaking the bipartiteness, does not change the results in any fundamental way. We presume that this is the case also when the mean fields are solved self-consistently.

We have also neglected the orbital effect of the magnetic field perpendicular to the quasicrystal. In this work, we have focused on the kinetic-energy range near the top or bottom of the band, where the dispersion in Penrose and AB QCs can be regarded quadratic as in a square lattice \cite{Ghadimi_2021}. Assuming a quadratic dispersion, the effective mass and hence a uniform magnetic field that corresponds to the Zeeman energy $h/t$ can be estimated. By incorporating the magnetic field in terms of the Peierls factor \cite{Smith_2016} and using OBC, we have checked for some of the systems studied above that even though it tends to require more iterations, the converged mean fields are very similar to those without the Peierls factor.
Moreover, we have confirmed the appearance of Majorana zero modes at the vortex centre and along the edges in the system presented in Fig.~\ref{fig:AB8119_vortex_KTsystems_vortexModes} with the Peierls factor included, with the excitation energy $\epsilon\simeq 4\times 10^{-5}t$.

In contrast to the kinetic-energy range that we have focused on in this work, the band structure can be very different from that of a square lattice in other energy ranges, e.g., due to possible formation of a pseudogap \cite{Zijlstra_2000PRB,Collins_2017,Araujo_2019} and the presence of strictly localised states \cite{Kohmoto_1986,Arai_1988,Koga_2020} in both QCs and with insulating gaps in Penrose QCs \cite{Kohmoto_1986PRL,Koga_2017}.
It is thus interesting to explore different regions of the kinetic-energy band for the possible existence of TSC \cite{Saito_2024}.
Momentum-space analysis of the band structure in terms of the spectral function \cite{Araujo_2019,Ghadimi_2021,Hori_2023} will be useful for understanding the nature of TSC and topological phase transitions.
Such studies are currently under way.

As an experimental setup for creating a two-dimensional quasicrystalline topological superconductor, we propose a Pb monolayer with inherent $s$-wave superconductivity and strong Rashba spin-orbit coupling deposited \cite{Ledieu_2008} on the fivefold-symmetric surface of a three-dimensional quasicrystal such as Al-Pd-Mn \cite{Ledieu_2004,Ledieu_2008,Smerdon_2008} or Al-Cu-Fe \cite{Sharma_2005}, with a single-layer of Co \cite{Menard_2017} deposited on top for generating the Zeeman field. Moreover, superconductivity has recently been observed in a van der Waals-layered Ta$_{1.6}$Te quasicrystal \cite{Tokumoto_23}, which consists of layered two-dimensional dodecagonal quasicrystals \cite{Conrad_1998}, with evidence of strong spin-orbit coupling \cite{Terashima_2024}. Therefore, Ta$_{1.6}$Te quasicrystals doped with magnetic impurities or layered with a ferromagnetic insulator are potential candidate materials for realising TSC with broken time-reversal symmetry.

\begin{acknowledgments}
We thank Rasoul Ghadimi, Aksel Kobia{\l}ka, Yoshi Maeno, Yuki Nagai, Ryo Okugawa, and Andrzej Ptok for stimulating discussions. K.T. is grateful to the Pauli Center for Theoretical Studies for support and the University of Z\"urich for support and hospitality, where part of the work was performed.
This work was supported by the Natural Sciences and Engineering Research Council of Canada, JST SPRING (Grant No. JPMJSP2151), and the Japan Society for the Promotion of Science, KAKENHI (Grant No. JP19H05821). 
We also acknowledge support by the Tokyo University of Science Grant for International Joint Research. 
This research was made feasible in part by support provided by Calcul Qu\'ebec, Compute Ontario, and the Digital Research Alliance of Canada (\href{https://alliancecan.ca/en}{alliancecan.ca}).
\end{acknowledgments}

\appendix
%\section{Square lattice}\label{appendix}
\section{Topological phase transitions in periodic systems}\label{appendix}

\begin{figure}[tbh]
  \includegraphics[width=9cm]{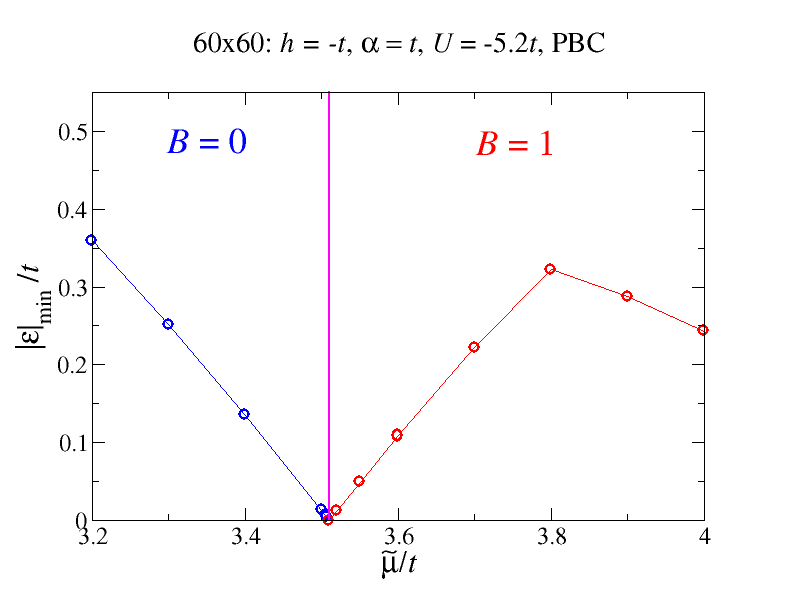}
\caption{Lowest absolute energy eigenvalue $|\epsilon|_{{\rm{min}}}/t$ as a function of the chemical potential $\tilde{\mu}/t$ in a $60\times 60$ square lattice for $h=-t$, $\alpha=t$, $U=-5.2t$, and PBC.
\label{fig:sqBott_mutilde}}
\end{figure}

\begin{figure}[tbh]
  \includegraphics[width=9cm]{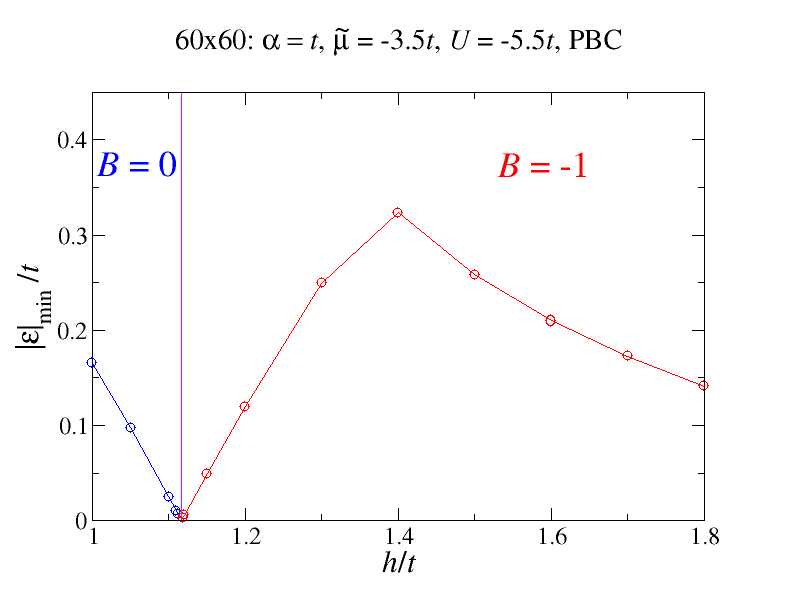}
\caption{Lowest absolute energy eigenvalue $|\epsilon|_{{\rm{min}}}/t$ as a function of the Zeeman energy $h/t$ in a $60\times 60$ square lattice for $\alpha=t$, $\tilde{\mu} = -3.5t$, $U=-5.5t$, and PBC.
\label{fig:sqBott_h}}
\end{figure}

Here we illustrate topological phase transitions in periodic systems which are analogous to those presented in Figs.~\ref{fig:PenBott_mutilde} and \ref{fig:PenBott_h} for $3571$-site Penrose QCs. In Fig.~\ref{fig:sqBott_mutilde} the lowest absolute excitation energy $|\epsilon|_{{\rm{min}}}$ is plotted as a function of $\tilde{\mu}/t$ for a $60\times 60$ square lattice with PBC for $h=-t$, $\alpha=t$, and $U=-5.2t$. The converged order parameter $\Delta_0$, which is uniform and real, decreases monotonically from $0.86t$ at $\tilde{\mu}=3.2t$ to $0.32t$ at $\tilde{\mu}=4t$ (not shown). 
At $\tilde{\mu} \simeq 3.51t$, $\Delta_0\simeq 0.61t$ and the effective Zeeman energy $\tilde{h}\simeq -0.78t$, satisfying the gap closing condition for $\tilde{\mu}>2t$ \cite{Sato_2010},
%\begin{equation}
$(4t-\tilde{\mu})^2 + \Delta_0^2 = \tilde{h}^2$,
%\end{equation}
and the TKNN number \cite{Thouless_1982} changes from 0 to $-1$. The Bott index ($B$) as defined in Eq.~(\ref{eq:Bott_index}) has the opposite sign and thus changes from 0 to 1. %at $\tilde{\mu} \simeq 3.51t$.

Shown in Fig.~\ref{fig:sqBott_h} is the absolute minimum eigenvalue $|\epsilon|_{{\rm{min}}}$ as a function of the Zeeman energy $h/t$ for a $60\times 60$ square lattice with PBC for $\alpha=t$, $\tilde{\mu}/t=-3.5t$, and $U=-5.5t$. From $h=t$ to $h=1.8t$, $\Delta_0$ ($\tilde{h}$) decreases (increases) monotonically from $0.85t$ ($0.82t$) to $0.22t$ ($1.33t$).
At $h\simeq 1.115t$, $\tilde{h}\simeq 0.88t$ and $\Delta_0\simeq 0.72t$, where the condition for $\tilde{\mu}\le -2t$ for a topological phase transition \cite{Sato_2010},
%\begin{equation}
$(4t+\tilde{\mu})^2 + \Delta_0^2 = \tilde{h}^2$, is satisfied and the TKNN number ($B$) changes from 0 to 1 ($-1$). 
%\end{equation}

%\appendix*

%\input sections/appendix1.tex

% \nocite{*}

\bibliography{bibs}% Produces the bibliography via BibTeX.

\end{document}